\documentclass [11pt,a4paper] {article}

\usepackage[T1]{fontenc}
\usepackage{lmodern}

\usepackage[cp1252]{inputenc}
\usepackage{amssymb}
\usepackage{amsmath}
\usepackage{amsfonts,amssymb}
\usepackage[dvips]{graphicx}
\usepackage{bbm}
\usepackage{enumerate}

\usepackage[shortlabels]{enumitem}

\usepackage{amsthm}
\usepackage{cancel}

\DeclareMathAlphabet{\mathpzc}{OT1}{pzc}{m}{it}

\def\SmallColSep{\setlength{\arraycolsep}{1pt}}

\newcommand*\rfrac[2]{{}^{#1}\!\!/\!_{#2}}

\begin{document}

\title{Overturning negative construal of quantum superposition}

\author{Arkady Bolotin\footnote{$Email: arkadyv@bgu.ac.il$\vspace{5pt}} \\ \emph{Ben-Gurion University of the Negev, Beersheba (Israel)}}

\maketitle

\begin{abstract}\noindent Construal of observable facts or events, that is, the manner in which we understand reality, is based not only on mathematical formulas of a theory suggested as a reasonable explanation for physical phenomena (like general relativity or quantum mechanics), but also on a mathematical model of reasoning used to analyze and appraise statements regarding the objective world (for example, logic of one type or the other). Hence, every time that a certain construal of reality encounters a problem, there is a choice between a modification to the mathematical formalism of the physical theory and a change in the model of reasoning. A case in point is negative construal of quantum superposition causing the problem of definite outcomes. To be sure, according to the said construal, it is not the case that a system being in a superposition of states is exclusively in one of the states constituting the superposition, which in turn implies that macroscopically differing outcomes of observation may appear all at once. The usual approach to the problem of definite outcomes is to modify the quantum mathematical formalism by adding to it some extra postulates (for instance, the postulate of wave function collapse). However, since none of the extra postulates proposed so far has gained broad acceptance, one may try another avenue to resolve the problem, namely, to replace logic with an alternative mathematical model of reasoning. This possibility is studied in the present paper.\\\\

\end{abstract}

\section{Introduction}  %{<-------------------------------------------------------------------------------------------------Section I}

\noindent Although the quantum measurement problem has been known for a long time, it is still difficult to find its exact definition in the literature. Besides, there happen to be not one, but several distinct issues collectively termed by this name \cite{Schlosshauer}. One of them is \emph{the problem of definite outcomes} (also known as \emph{the “and/or” problem}) \cite{Despagnat}. It asks how a single actual outcome of measurement can be reconciled with two or more potential outcomes generated by the mathematical formalism of quantum mechanics. The difficulty in bringing into agreement actual and potential outcomes is that \emph{the instrumentalist description of quantum mechanics} (i.e., the description which merely relates the mathematical formalism of quantum theory to data and prediction) does not contain a semantic explanation of \emph{a superposition of states} \cite{Timpson}.\\

\noindent In a bit more detail, the instrumentalist description of quantum mechanics associates a state of a physical system and a value of an observable in this state (corresponding to the outcome of measurement therein) with, respectively, a Hilbert space vector and a Hermitian operator acting on this vector. If the said vector is presented as a linear combination of other vectors of the Hilbert space, then the instrumentalist description simply declares that the system is in a superposition of states. For all that, an explanation of what this superposition might mean is left out.\\

\noindent According to \emph{negative construal of quantum superposition} established by general assent (see the references \cite{Leggett07, Reid} to cite but a few), the system in a superposition of theoretically possible states represents a situation where it is not the case that the system is exclusively in one of the possible states. A problem arises when this way of understanding quantum superposition is applied to a macroscopic object: In that case it necessitates the conclusion that “macroscopically differing” contradictory quantities (like different pointer positions of a measurement apparatus) may appear all at once. However, suffice to say that in a typical Young’s double-slit experiment, any effective attempt to measure which path an individual particle took yields a definite pointer position, that is, each particle appears to take either the path through one slit or that through the other.\\

\noindent So, unless negative construal of superposition were to be overturned or an extra postulate were to be supplemented to the instrumentalist description of quantum mechanics, the premise of either “\emph{quantum fundamentalism}” (asserting that everything in the universe is fundamentally of a quantum nature and thus describable in quantum-mechanical terms \cite{Zinkernagel}) or “\emph{macroscopic realism}” (declaring that in a situation involving macroscopically distinct contradictory outcomes of measurement, the occurrence of the one precludes the occurrence of the others \cite{Leggett85}) would be false.\\

\noindent The usual line of attack to the problem of definite outcomes is to come out with some extra postulate (or postulates). As an example, in the Copenhagen interpretation of quantum mechanics (commonly named the ``standard'' interpretation \cite{Faye}), the extra postulate is that the measurement process randomly picks out exactly one of the potential outcomes allowed for by the mathematical formalism of quantum mechanics (in a manner consistent with the well-defined probabilities that are assigned to each potential outcome) \cite{Neumann}. As a result, the act of measuring (or the act of observing if a numerical value is attached to the observed phenomenon by counting or measuring) causes all the potential outcomes to fall into one actual outcome (this process is known as wave function collapse).\\

\enlargethispage{\baselineskip}
\enlargethispage{\baselineskip}
\noindent Without regard to specific details or exceptions, it is safe to say that a multiplicity of current interpretations of quantum mechanics is a result of a great variety of proposed supplemental postulates. Then again, none of the existing interpretations of quantum mechanics is settled by common consent \cite{Vaidman}. A possible explanation for this fact is that some of the existing interpretations are \emph{unfalsifiable}, ergo, uncriticizable (no evidence that comes to light can contradict them and therefore they cannot be empirically assessed), while the others render themselves \emph{unlikely} by making predictions, which are different from those of the instrumentalist description of quantum mechanics. The situation with interpretations of quantum mechanics casts doubt on the viability of the steps commonly taken to resolve the problem of definite outcomes, that is, adding more postulates to the quantum instrumentalist description.\\

\noindent Accepting that this description needs no extra postulate to explain quantum phenomena compels one to seek ways to overturn negative construal of quantum superposition.\\

\noindent Such ways will be examined in the present paper.\\

\section{Components of negative construal of quantum superposition}  %{<-------------------------------------------------------------------------------------------------Section II}

\noindent Let us start by analyzing components of negative construal of quantum superposition, namely, concepts, hypotheses, and assumptions on which this construal is premised.\\

\noindent To make our analysis tangible, let us turn to \emph{an ideal coin}, explicitly, an idealized macroscopic randomizing device with just two states. It can be visualized as a macroscopic two-sided object (akin to a flat circular plate) that cannot be set on its edge because it has an almost zero thickness. Imagine tossing this coin repetitively and observing on every toss which side – “heads” or “tails” – is up when it lands on a horizontal surface.\\

\noindent Let us introduce into our analysis the concept of \emph{categorical property}. Recall that a categorical property is a characteristic of an object that concerns what quality the object has. By contrast, a dispositional property involves what the object can do \cite{Prior, Strawson}. For example, a side showing up after the landing is a categorical property of the coin, while the electrical conductivity of the coin is its dispositional property.\\

\noindent Consider the statement \text{\guillemotleft}The categorical property of the coin is “heads up”\text{\guillemotright} (here and hereinafter in this paper, the double angle quotation marks are used to set statements off from the rest of the text). Let this statement be denoted by the letter $H$ for brevity. Similarly, the statement \text{\guillemotleft}The categorical property of the coin is “tails up”\text{\guillemotright} is denoted by $T$.\\

\begin{figure}[ht!]
   \centering
   \includegraphics[scale=0.5]{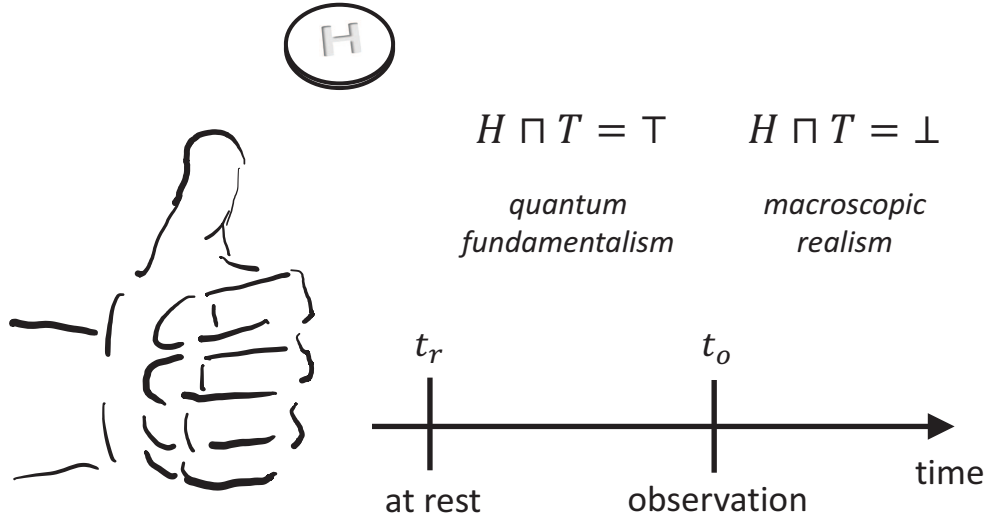}
   \caption{Quantum fundamentalism vs. macroscopic realism in coin tossing.\label{fig1}}
\end{figure}

\noindent Let $t_r$ signify the point in time $t$ at which the coin comes to rest after it lands (see Figure \ref{fig1}). Understandably, the statements $H$ and $T$ may have no definite truth value until that point. Assume that from $t_r$ onward, the statements $H$ and $T$ obey \emph{the principle of bivalence}, namely, each of them is either true or false. Because the ideal coin cannot land on its edge and stay there, it cannot rest on a horizontal surface showing up neither “heads” nor “tails”. Consequently, after the moment $t_r$ the statements $H$ and $T$ are not able to be false together.\\

\noindent Let us stipulate that the adverb “not” expresses negation, which changes an affirmative statement into a negative one, and vice versa. Then, using the logical connectives $\neg$ and $\sqcap$ that represent negation and conjunction, respectively, \emph{the requirement for idealness of the coin} can be written down as\smallskip

\begin{equation}  %{Eq.1}
   t
   \in
   [t_r,+\infty)
   \mkern-3.3mu
   :
   \mkern4mu
   \neg
   H
   \sqcap
   \neg
   T
   \mkern-2mu
   \iff
   \bot
   \;\;\;\;  ,
\end{equation}
\smallskip

\noindent where the logical connective $\iff$ corresponds to the expression ``is equivalent to'' and $\bot$ denotes an arbitrary contradiction. Providing De Morgan’s laws and double negative elimination hold, one gets\smallskip

\begin{equation}  %{Eq.2}
   t
   \in
   [t_r,+\infty)
   \mkern-3.3mu
   :
   \mkern4mu
   H
   \sqcup
   T
   \mkern-2mu
   \iff
   \top
   \;\;\;\;  ,
\end{equation}
\smallskip

\noindent where $\top\iff\neg\bot$ denotes an arbitrary tautology.\\

\noindent Let $t_o$, where $t_o{\mkern-0.5mu >\mkern2mu}t_r$, be the moment at which the coin is observed. In accordance with the hypothesis of “macroscopic realism”, “heads” and “tails” cannot be seen simultaneously. This means that after $t_o$ the statements $H$ and $T$ are not able to be true together. Presented in logical terms, this may be written directly as\smallskip

\begin{equation}  %{Eq.3}
   t
   \in
   [t_o,+\infty)
   \mkern-3.3mu
   :
   \mkern4mu
   \neg
   H
   \sqcup
   \neg
   T
   \mkern-2mu
   \iff
   \top
   \;\;\;\;  .
\end{equation}
\smallskip

\noindent Hence, using De Morgan’s laws and double negative elimination, \emph{the requirement of “macroscopic realism”} can be recorded as follows:\smallskip

\begin{equation} \label{AFTER} %{Eq.4}
   t
   \in
   [t_o,+\infty)
   \mkern-3.3mu
   :
   \mkern4mu
   H
   \sqcap
   T
   \mkern-2mu
   \iff
   \bot
   \;\;\;\;  .
\end{equation}
\smallskip

\noindent Now, consider a moment preceding $t_o$ but following $t_r$, explicitly, $t_r{\mkern2mu\le\mkern2mu}t{\mkern-0.5mu <\mkern2mu}t_o$. What is the truth value of the compound statement $H\mkern2mu{\sqcap}\mkern2muT$ at such a moment?\\

\noindent If the hypothesis of “quantum fundamentalism” holds true, then during the interval $[t_r,t_o)$, the coin is in a state formed by a superposition of two macroscopically distinct states, namely, the state where “heads” shows up and the state where “tails” shows up. If quantum superposition is construed in a negative sense, then the state of the coin at $t\in[t_r,t_o)$ corresponds to a situation where \emph{it is not the case that the statements $H$ and $T$ are not able to be true together}. Stipulating that the expression “it is not the case” is equivalent to negation, the aforesaid correspondence follows the logic that the negation of ${\neg}H{\sqcup}{\neg}T$ must be true on every toss of the coin, i.e.,\smallskip

\begin{equation}  %{Eq.5}
   t
   \in
   [t_r,t_o)
   \mkern-3.3mu
   :
   \mkern4mu
   \neg
   \left(
      \neg
      H
      \sqcup
      \neg
      T
   \right)
   \mkern-2mu
   \iff
   \top
   \;\;\;\;  .
\end{equation}
\smallskip

\noindent Using De Morgan’s laws and double negative elimination once more, this is translated into the following sentence:\smallskip

\begin{equation} \label{PRIOR} %{Eq.6}
   t
   \in
   [t_r,t_o)
   \mkern-3.3mu
   :
   \mkern4mu
   H
   \sqcap
   T
   \mkern-2mu
   \iff
   \top
   \;\;\;\;  .
\end{equation}
\smallskip

\noindent Combining (\ref{PRIOR}) with (\ref{AFTER}), the compound statement $H{\mkern1.5mu\sqcap\mkern1.5mu}T$ can be presented as a propositional (statemental) function of time $t$ (this function becomes a statement when $t$ takes on a definite value). Explicitly,\smallskip

\begin{equation}  %{Eq.7}
   \left(
      H
      \sqcap
      T
   \right)_{t}
   \mkern-2mu
   \iff
   \left\{
      \begingroup\SmallColSep
      \begin{array}{r l}
         \top
         &
         \mkern3mu
         ,
         \mkern12mu
         t
         \in
         [t_r,t_o)
         \\[5pt]
         \bot
         &
         \mkern3mu
         ,
         \mkern12mu
         t
         \in
         [t_o,+\infty)
      \end{array}
      \endgroup   
   \right.   
   \;\;\;\;  .
\end{equation}
\smallskip

\noindent As it is seen, under negative construal of quantum superposition, a reconciliation between “quantum fundamentalism” and “macroscopic realism” struggles against the fact that the limit of the function $\left(H\mkern2mu{\sqcap}\mkern2muT\right)_{t}$ as $t$ approaches $t_o$ from the left does not exist. In consequence, the resolution of the problem of definite outcomes turns into the removal of the jump discontinuity of this function at the instant $t_o$ – something that cannot be done without an extra postulate (or postulates).\\

\noindent For example, translated into logical terms, the postulate of wave function collapse says that the act of observing causes a tautology $\top$ to change into a contradiction $\bot$ at the moment $t_o$. However, changes like this could only be possible if the truth values “true” and “false” were not objective (absolute) but relative to experience (e.g., an observation or experiment). Hence, accepting the ``standard'' interpretation of quantum mechanics amounts to giving up the objectivity of “true” and “false”!\\

\noindent Hypothetically, “true” and “false” would not be influenced or controlled by observation if the compound statement $H\mkern2mu{\sqcap}\mkern2muT$ were to have the same truth value over the whole interval $[t_r,+\infty)$. But in order to convert $\left(H\mkern2mu{\sqcap}\mkern2muT\right)_{t}$ into a constant function without breaking down either “quantum fundamentalism” or “macroscopic realism”, negative construal of quantum superposition must be thrown out.\\

\noindent Let us consider how to accomplish this.\\

\section{Requirements on negative construal of quantum superposition}  %{<-------------------------------------------------------------------------------------------------Section III}

\noindent Besides De Morgan’s laws, the conditions, under which negative construal of quantum superposition can be accepted, are as follows:\\[-15pt]

\begin{enumerate}[(a)]
    \item a truth value of a compound statement entirely depends on truth values of the statements that make it up,
    \item a truth value of a negated statement is contradictory to the truth value of the statement itself,
    \item a relation between the set of statements and the set of the truth values is a function,
    \item this function is total,
    \item the set of the truth values has just two members,
    \item these members are “true” and “false”.
\end{enumerate}

\noindent According to the conditions (c)-(f), the relation between the set of statements, denoted $\mathbb{S}$, and the set of the truth values, denoted $\mathbb{B}_2$, is a total surjective-only function $v:\mkern2mu\mathbb{S}\to\mathbb{B}_2$, where $|\mathbb{B}_2|$, the cardinality of $\mathbb{B}_2$, is 2; explicitly, $\mathbb{B}_2 =\{\text{``true''},\text{``false''}\}$ or $\mathbb{B}_2 =\{\text{T},\text{F}\}$ or $\mathbb{B}_2 =\{1,0\}$. The image of a statement, say, $S$, under $v$ is denoted by $[\mkern-3.3mu[S]\mkern-3.3mu]=v(S)$. Consequently, the relation between $S$ and $\mathbb{B}_2$ can be presented in the arrow notation as\smallskip

\begin{equation} \label{TV} %{Eq.8}
   \begingroup
   \begin{array}{r c c c}
      v:
      &
      \mathbb{S}
      &
      \to
      &
      \mathbb{B}_2
      \\[5pt]
      \hfill
      &
      S
      &
      \mapsto
      &
      [\mkern-3.3mu[S]\mkern-3.3mu]
   \end{array}
   \endgroup
   \;\;\;\;  ,
\end{equation}
\smallskip

\noindent where the first part is read “$v$ is a total function from $\mathbb{S}$ to $\mathbb{B}_2$”, while the second part is read “$S$, an element of $\mathbb{S}$, maps onto $[\mkern-3.3mu[S]\mkern-3.3mu]$, an element of $\mathbb{B}_2$”.\\

\noindent Consider a statement which is unable to be broken down into smaller statements. Due to this reason, it is called \emph{an atomic statement} or simply \emph{an ``atom''}. Let us denote a compound statement made from two ``atoms'', say, $S_1$ and $S_2$, by $S_1{\ast}S_2$, where $\ast$ stands for a logical (binary) connective (such as $\sqcap$, $\sqcup$ and so on). Then, in the symbolic form, the condition (a) can be written as\smallskip

\begin{equation} \label{BCON} %{Eq.9}
   [\mkern-3.3mu[
      S_1{\ast}S_2
   ]\mkern-3.3mu]
   \mkern-2mu
   =
   f_{\mkern-2mu\ast}
   \mkern-2mu
   \left(
      [\mkern-3.3mu[S_1]\mkern-3.3mu]
      ,
      [\mkern-3.3mu[S_2]\mkern-3.3mu]
   \right)
   \;\;\;\;  ,
\end{equation}
\smallskip

\noindent where $f_{\mkern-2mu\ast}$ represents a total (surjective-only) function associated with the particular binary connective $\ast$. For example, consistent with the basic operations of Boolean calculus \cite{Hailperin}, $f_{\mkern-2mu\sqcap}([\mkern-3.3mu[S_1]\mkern-3.3mu],[\mkern-3.3mu[S_2]\mkern-3.3mu])=\min([\mkern-3.3mu[S_1]\mkern-3.3mu],[\mkern-3.3mu[S_2]\mkern-3.3mu])$ and $f_{\mkern-2mu\sqcup}([\mkern-3.3mu[S_1]\mkern-3.3mu],[\mkern-3.3mu[S_2]\mkern-3.3mu])=\max([\mkern-3.3mu[S_1]\mkern-3.3mu],[\mkern-3.3mu[S_2]\mkern-3.3mu])$.\\

\noindent Using (\ref{BCON}) as an exemplar, the condition (b) can be given by\smallskip

\begin{equation}  %{Eq.10}
   [\mkern-3.3mu[
      {\neg}S
   ]\mkern-3.3mu]
   \mkern-2mu
   =
   f_{\mkern-2mu\neg}
   \mkern-2mu
   \left(
      [\mkern-3.3mu[S]\mkern-3.3mu]
   \right)
   =
   1
   -
   [\mkern-3.3mu[S]\mkern-3.3mu]
   \;\;\;\;  .
\end{equation}
\smallskip

\noindent As it follows, to overturn negative construal of quantum superposition, one has to modify (replace or discard) at least one of the expressions (\ref{TV}) and (\ref{BCON}).\\

\section{Proposals to modify truth-functionality of logic connectives}  %{<-------------------------------------------------------------------------------------------------Section IV}

\noindent Apparently, the “weakest” among these two is (\ref{BCON}), the expression that represents the principle underlying bivalent truth-functional propositional (statemental) logic (usually called \emph{classical logic}) \cite{Klement}.\\

\noindent Indeed, let us recall that in accordance with the instrumentalist description of quantum mechanics, a truth value of the statement $S$ is given by either eigenvalue of the observable $\hat{O}_{S}$, i.e., a Hermitian operator whose spectrum is the two-element set $\{0,1\}$ (it is said that the observable $\hat{O}_{S}$ encodes the statement $S$). On the other hand, two observables $\hat{O}_{S_1}$ and $\hat{O}_{S_2}$ have a simultaneous set of eigenvectors if and only if the observables commute. Providing the presence or the lack of commutativity of $\hat{O}_{S_1}$ and $\hat{O}_{S_2}$ is measured by their commutator defined by\smallskip

\begin{equation}  %{Eq.11}
   \left[
      \hat{O}_{S_1}
      ,
      \hat{O}_{S_2}
   \right]
   \mkern-2mu
   =
      \hat{O}_{S_1}
      \hat{O}_{S_2}
      -
      \hat{O}_{S_2}
      \hat{O}_{S_1}
   \;\;\;\;  ,
\end{equation}
\smallskip

\noindent the observables commute if their commutator is the zero operator $\hat{0}$ (which takes any vector $|\Psi\rangle$ of a Hilbert space to the zero vector, 0, belonging to the zero subspace $\{0\}$; the zero operator encodes an arbitrary contradiction $\bot$, so that the observable $\hat{O}_{\bot}$ is $\hat{0}$). Contrastively, if $[\hat{O}_{S_1},\hat{O}_{S_2}]\neq\hat{0}$, then the statements $S_1$ and $S_2$, which are encoded by the observables $\hat{O}_{S_1}$ and $\hat{O}_{S_2}$, cannot have truth values at the same time. For this reason, the basic Boolean functions, e.g.,\smallskip

\begin{equation}  %{Eq.12}
   [\mkern-3.3mu[
      S_1{\sqcap}S_2
   ]\mkern-3.3mu]
   \mkern-2mu
   =
   \min
   \mkern-2mu
   \left(
      [\mkern-3.3mu[S_1]\mkern-3.3mu]
      ,
      [\mkern-3.3mu[S_2]\mkern-3.3mu]
   \right)
   \;\;\;\;  ,
\end{equation}
\\[-36pt]

\begin{equation}  %{Eq.13}
   [\mkern-3.3mu[
      S_1{\sqcup}S_2
   ]\mkern-3.3mu]
   \mkern-2mu
   =
   \max
   \mkern-2mu
   \left(
      [\mkern-3.3mu[S_1]\mkern-3.3mu]
      ,
      [\mkern-3.3mu[S_2]\mkern-3.3mu]
   \right)
   \;\;\;\;  ,
\end{equation}
\smallskip

\noindent cannot be defined properly when $[\hat{O}_{S_1},\hat{O}_{S_2}]\neq\hat{0}$.\\

\noindent To address uncommunicativeness of observables, one may propose to modify classical logic in such a way that the above formulas take the form:\smallskip

\begin{equation} \label{QCON} %{Eq.14}
   [\mkern-3.3mu[
      S_1{\sqcap}S_2
   ]\mkern-3.3mu]
   \mkern-2mu
   =
   \left\{
      \begingroup\SmallColSep
      \begin{array}{l l}
         \min
         \mkern-2mu
         \left(
            [\mkern-3.3mu[S_1]\mkern-3.3mu]
            ,
            [\mkern-3.3mu[S_2]\mkern-3.3mu]
         \right)
         &
         \mkern3mu
         ,
         \mkern12mu
         \left[
            \hat{O}_{S_1}
            ,
            \hat{O}_{S_2}
         \right]
         \mkern-2mu
         =
         \hat{0}
         \\[5pt]
         0
         &
         \mkern3mu
         ,
         \mkern12mu
         \left[
            \hat{O}_{S_1}
            ,
            \hat{O}_{S_2}
         \right]
         \mkern-2mu
         \neq
         \hat{0}
      \end{array}
      \endgroup   
   \right.   
   \;\;\;\;  ,
\end{equation}
\\[-20pt]

\begin{equation} \label{QDIS} %{Eq.15}
   [\mkern-3.3mu[
      S_1{\sqcup}S_2
   ]\mkern-3.3mu]
   \mkern-2mu
   =
   \left\{
      \begingroup\SmallColSep
      \begin{array}{l l}
         \max
         \mkern-2mu
         \left(
            [\mkern-3.3mu[S_1]\mkern-3.3mu]
            ,
            [\mkern-3.3mu[S_2]\mkern-3.3mu]
         \right)
         &
         \mkern3mu
         ,
         \mkern12mu
         \left[
            \hat{O}_{S_1}
            ,
            \hat{O}_{S_2}
         \right]
         \mkern-2mu
         =
         \hat{0}
         \\[5pt]
         1
         &
         \mkern3mu
         ,
         \mkern12mu
         \left[
            \hat{O}_{S_1}
            ,
            \hat{O}_{S_2}
         \right]
         \mkern-2mu
         \neq
         \hat{0}
      \end{array}
      \endgroup   
   \right.   
   \;\;\;\;  .
\end{equation}
\smallskip

\noindent It is easy to demonstrate that this proposal (originally presented in \cite{Birkhoff}) implies a breakdown of the distributive law for the logical operators $\sqcap$ and $\sqcup$ (and so it can be called \emph{a proposal of a non-distributive propositional logic}).\\

\noindent Providing the observable $\hat{O}_{{\neg}S}$ that encodes ${\neg}S$ is given by\smallskip

\begin{equation}  %{Eq.16}
   \hat{O}_{{\neg}S}
   =
   \hat{1}
   -
   \hat{O}_{S}
   \;\;\;\;  ,
\end{equation}
\smallskip

\noindent where $\hat{1}$ represents the identity operator (which takes any vector $|\Psi\rangle$ to itself and encodes an arbitrary tautology $\top$; consequently, $\hat{O}_{\top}=\hat{1}$), one gets $[\hat{O}_{S},\hat{O}_{{\neg}S}]=\hat{0}$. Using (\ref{QCON}) and (\ref{QDIS}) one then realizes that the compound statements $S\sqcap{\neg}S$ and $S\sqcup{\neg}S$ are a contradiction and a tautology, respectively,\smallskip

\begin{equation}  %{Eq.17}
   \left[
      \hat{O}_{S}
       ,
      \hat{O}_{{\neg}S}
   \right]
   \mkern-2mu
   =
   \hat{0}
   :
   \mkern12mu
       \begingroup\SmallColSep
      \begin{array}{l}
         [\mkern-3.3mu[S\sqcap{\neg}S]\mkern-3.3mu]         
         =
         \min
         \mkern-2mu
         \left(
            [\mkern-3.3mu[S]\mkern-3.3mu]
            ,
            1
            -
            [\mkern-3.3mu[S]\mkern-3.3mu]
         \right)
         =
         0
         \\[5pt]
         [\mkern-3.3mu[S\sqcup{\neg}S]\mkern-3.3mu]         
         =
         \max
         \mkern-2mu
         \left(
            [\mkern-3.3mu[S]\mkern-3.3mu]
            ,
            1
            -
            [\mkern-3.3mu[S]\mkern-3.3mu]
         \right)
         =
         1
      \end{array}
      \endgroup   
   \;\;\;\;  .
\end{equation}
\smallskip

\noindent At this point, consider the expression arising from the distributivity law:\smallskip

\begin{equation} \label{DIST} %{Eq.18}
   S_2
   \sqcup
   \left(
      S_1
      \sqcap
      {\neg}S_1
   \right)
   \iff
   \left(
      S_2
      \sqcup
      S_1
   \right)
   \sqcap
   \left(
      S_2
      \sqcup
      {\neg}S_1
   \right)   
   \;\;\;\;  .
\end{equation}
\smallskip

\noindent Since the observable $\hat{O}_{S_2}$ commutes with the observable $\hat{O}_{\bot}$ encoding the contradiction $S_1\sqcap{\neg}S_1$, a truth value of the left-hand side of the above expression is\smallskip

\begin{equation}  %{Eq.19}
   \left[
      \hat{O}_{S_2}
       ,
      \hat{O}_{\bot}
   \right]
   \mkern-2mu
   =
   \hat{0}
   :
   \mkern12mu
   [\mkern-3.3mu[
      S_2
      \sqcup
      \left(
         S_1
         \sqcap
         {\neg}S_1
      \right)
   ]\mkern-3.3mu]         
   =
   \max
   \mkern-2mu
   \left(
      [\mkern-3.3mu[S_2]\mkern-3.3mu]
      ,
      0
   \right)
   =
   [\mkern-3.3mu[S_2]\mkern-3.3mu]
   \;\;\;\;  .
\end{equation}
\smallskip

\noindent Suppose $[\hat{O}_{S_1},\hat{O}_{S_2}]\neq\hat{0}$. Therefore, $[\hat{O}_{{\neg}S_1},\hat{O}_{S_2}]\neq\hat{0}$. Then, in accordance with (\ref{QDIS}), $S_2{\mkern1.5mu\sqcup\mkern1.5mu}S_1$ and $S_2{\mkern1.5mu\sqcup\mkern1.5mu}{\neg}S_1$ are tautologies. As the observable $\hat{O}_{\top}$ commutes with itself, a truth value of the right-hand side of the expression (\ref{DIST}) comes to be\smallskip

\begin{equation}  %{Eq.20}
   \left[
      \hat{O}_{\top}
       ,
      \hat{O}_{\top}
   \right]
   \mkern-2mu
   =
   \hat{0}
   :
   \mkern12mu
   [\mkern-3.3mu[
      \left(
         S_2
         \sqcup
        S_1
      \right)
      \sqcap
      \left(
         S_2
         \sqcup
        {\neg}S_1
      \right)
   ]\mkern-3.3mu]
   =
   \min
   \mkern-2mu
   \left(
      [\mkern-3.3mu[S_2{\mkern2mu\sqcup\mkern2mu}S_1]\mkern-3.3mu]
      ,
      [\mkern-3.3mu[S_2{\mkern2mu\sqcup\mkern2mu}{\neg}S_1]\mkern-3.3mu]
   \right)
   =
   1
   \;\;\;\;  .
\end{equation}
\smallskip

\noindent Given that an arbitrary statement $S_2$ is contingent and thus not a tautology, one concludes that the expression (\ref{DIST}) does not hold true for the most part of occasions.\\

\noindent The failure of distributivity of $\sqcup$ over $\sqcap$ can be taken as a reason to overturn negative construal of quantum superposition.\\

\noindent To be sure, let us go back to the ideal coin. Consider the statement \text{\guillemotleft}The categorical property of the coin is “at rest”\text{\guillemotright} denoted by $R$. Correspondingly, ${\neg}R$ denotes the negation of this statement. The observable $\hat{O}_{R}$ encoding the statement $R$ and the observable $\hat{O}_{H}$ encoding the statement $H$ do not have the simultaneous set of eigenvectors. The same holds true for $\hat{O}_{R}$ and $\hat{O}_{T}$, the observable that encodes the statement $T$. Certainly, let $|\Psi_R\rangle$, $|\Psi_H\rangle$ and $|\Psi_T\rangle$ denote eigenvectors of the observables $\hat{O}_{R}$, $\hat{O}_{H}$ and $\hat{O}_{T}$. According to “quantum fundamentalism”, during the interval $[t_r,t_o)$, the coin is in the state $|\Psi_R\rangle$ formed by a superposition of two states,  $|\Psi_H\rangle$ and $|\Psi_T\rangle$; in symbols,\smallskip

\begin{equation}  %{Eq.21}
   |\Psi_R\rangle
   =
   c_H
   |\Psi_H\rangle
   +
   c_T
   |\Psi_T\rangle
   \;\;\;\;  ,
\end{equation}
\smallskip

\noindent where $c_H$ and $c_T$ are complex coefficients. As it can be seen, $|\Psi_R\rangle\mkern-2mu\notin\mkern-2mu\{c_H|\Psi_H\rangle\mkern-2mu|\mkern2mu c_H\mkern-2mu\in\mkern-2mu\mathbb{C},c_H\neq0\}$ and $|\Psi_R\rangle\mkern-2mu\notin\mkern-2mu\{c_T|\Psi_T\rangle\mkern-2mu|\mkern2mu c_T\mkern-2mu\in\mkern-2mu\mathbb{C},c_T\neq0\}$. This means\smallskip

\begin{equation}  %{Eq.22}
   {\forall}S_1\mkern-2mu\in\mkern-2mu\{R,{\neg}R\}
   ,
   \mkern4mu
    {\forall}S_2\mkern-2mu\in\mkern-2mu\{H,T\}
   :
   \mkern12mu
   \left[
      \hat{O}_{S_1}
       ,
      \hat{O}_{S_2}
   \right]
   \mkern-2mu
   \neq
   \hat{0}
   \;\;\;\;  .
\end{equation}
\smallskip

\noindent Hence, in accordance with (\ref{QDIS}), ${\neg}R{\mkern1.5mu\sqcup\mkern1.5mu}H$ and ${\neg}R{\mkern1.5mu\sqcup\mkern1.5mu}T$ are tautologies.\\

\noindent Now, consider the sentence ${\neg}R{\mkern1.5mu\sqcup\mkern1.5mu}(H{\mkern1.5mu\sqcap\mkern1.5mu}T)$. Since the negation of the statement $R$ becomes false after the coin comes to rest, the following holds when $t{\mkern1mu\ge\mkern2mu}t_r$:\smallskip

\begin{equation}  %{Eq.23}
   [\mkern-3.3mu[
      {\neg}R
      \sqcup
      \left(
         H
         \sqcap
         T
      \right)
   ]\mkern-3.3mu]         
   =
   \max
   \mkern-2mu
   \left(
      [\mkern-3.3mu[{\neg}R]\mkern-3.3mu]
      ,
      [\mkern-3.3mu[H{\mkern1.5mu\sqcap\mkern1.5mu}T]\mkern-3.3mu]
   \right)
   =
   [\mkern-3.3mu[H{\mkern1.5mu\sqcap\mkern1.5mu}T)]\mkern-3.3mu]
   \;\;\;\;  .
\end{equation}
\smallskip

\noindent On the other hand, had the distributive law for the logical operators $\sqcap$ and $\sqcup$ been in force during the period $[t_r,t_o)$, one would have obtained in addition:\smallskip

\begin{equation}  %{Eq.24}
   [\mkern-3.3mu[H{\mkern1.5mu\sqcap\mkern1.5mu}T)]\mkern-3.3mu]
   =
    [\mkern-3.3mu[
      \left(
         {\neg}R
         \sqcup
         H
      \right)
      \sqcap
      \left(
         {\neg}R
         \sqcup
         T
      \right)
   ]\mkern-3.3mu]
   =
   1
   \;\;\;\;  .
\end{equation}
\smallskip

\noindent But this law did not hold therein, which means\smallskip

\begin{equation}  %{Eq.25}
   [\mkern-3.3mu[H{\mkern1.5mu\sqcap\mkern1.5mu}T]\mkern-3.3mu]
   \neq
   1
   \;\;\;\;  .
\end{equation}
\smallskip

\noindent This indicates cancelation of negative construal of quantum superposition during the mentioned period.\\

\noindent Another way to tackle compound statements whose constituents are encoded by non-commutative observables is to propose that such statements are \emph{undefined} (where the precise meaning of the adjective “undefined” in the context of truth assignments is “not having been given a truth value”). According to this proposal, which can be called \emph{a proposal of partial propositional functions} (or \emph{partial predicates}) \cite{Specker, Kochen15}, the following holds true for the logical operators $\sqcup$ and $\sqcap$:\smallskip

\begin{equation}  %{Eq.26}
   [\mkern-3.3mu[
      S_1{\sqcap}S_2
   ]\mkern-3.3mu]
   \mkern-2mu
   =
   \left\{
      \begingroup\SmallColSep
      \begin{array}{l l}
         \min
         \mkern-2mu
         \left(
            [\mkern-3.3mu[S_1]\mkern-3.3mu]
            ,
            [\mkern-3.3mu[S_2]\mkern-3.3mu]
         \right)
         &
         \mkern3mu
         ,
         \mkern12mu
         \left[
            \hat{O}_{S_1}
            ,
            \hat{O}_{S_2}
         \right]
         \mkern-2mu
         =
         \hat{0}
         \\[5pt]
         \text{undefined}
         &
         \mkern3mu
         ,
         \mkern12mu
         \left[
            \hat{O}_{S_1}
            ,
            \hat{O}_{S_2}
         \right]
         \mkern-2mu
         \neq
         \hat{0}
      \end{array}
      \endgroup   
   \right.   
   \;\;\;\;  ,
\end{equation}
\\[-20pt]

\begin{equation}  %{Eq.27}
   [\mkern-3.3mu[
      S_1{\sqcup}S_2
   ]\mkern-3.3mu]
   \mkern-2mu
   =
   \left\{
      \begingroup\SmallColSep
      \begin{array}{l l}
         \max
         \mkern-2mu
         \left(
            [\mkern-3.3mu[S_1]\mkern-3.3mu]
            ,
            [\mkern-3.3mu[S_2]\mkern-3.3mu]
         \right)
         &
         \mkern3mu
         ,
         \mkern12mu
         \left[
            \hat{O}_{S_1}
            ,
            \hat{O}_{S_2}
         \right]
         \mkern-2mu
         =
         \hat{0}
         \\[5pt]
         \text{undefined}
         &
         \mkern3mu
         ,
         \mkern12mu
         \left[
            \hat{O}_{S_1}
            ,
            \hat{O}_{S_2}
         \right]
         \mkern-2mu
         \neq
         \hat{0}
      \end{array}
      \endgroup   
   \right.   
   \;\;\;\;  .
\end{equation}
\smallskip

\enlargethispage{\baselineskip}
\enlargethispage{\baselineskip}
\noindent Using the same line of reasoning that was applied to the proposal of a non-distributive propositional logic, one realizes that while $[\mkern-3.3mu[{\neg}R]\mkern-3.3mu]=0$, both $[\mkern-3.3mu[{\neg}R{\mkern1.5mu\sqcup\mkern1.5mu}H]\mkern-3.3mu]$ and $[\mkern-3.3mu[{\neg}R{\mkern1.5mu\sqcup\mkern1.5mu}T]\mkern-3.3mu]$ are undefined. This can be interpreted as a fact that throughout the interval $[t_r,t_o)$ neither $H$ nor $T$ can have a defined truth value simultaneously with ${\neg}R$. On the other hand, as long as the premise of “macroscopic realism” holds, $[\mkern-3.3mu[H{\mkern1.5mu\sqcap\mkern1.5mu}T]\mkern-3.3mu]=0$. Hence, the compound statement $H{\mkern1.5mu\sqcap\mkern1.5mu}T$ is false even if its constituents are undefined. What is more, seeing as both $[\mkern-3.3mu[H]\mkern-3.3mu]$ and $[\mkern-3.3mu[T]\mkern-3.3mu]$ are undefined when $[\mkern-3.3mu[{\neg}R]\mkern-3.3mu]=0$, negative construal of quantum superposition (saying that at $t\in[t_r,t_o)$ both $H$ and $T$ must only be true) can be dismissed.\\

\noindent Consequently, by modifying truth-functionality of the logical operators $\sqcap$ and $\sqcup$, “macroscopic realism” and “quantum fundamentalism” can hold simultaneously.\\

\noindent Still and all, the modifications of the operations $\sqcap$ and $\sqcup$ shown above cannot be accepted as an authentic solution to the problem of definite outcomes. This is so because of the extraneous nature of the commutator $[\hat{O}_{S_1},\hat{O}_{S_2}]$, on which the said modifications are contingent.\\

\noindent Let us delve into this point in a bit more detail. The commutator of two observables, $[\hat{O}_{S_1},\hat{O}_{S_2}]$, cannot be expressed by syntax of a propositional logic to the extent that the condition $[\hat{O}_{S_1},\hat{O}_{S_2}]=\hat{0}$ or $[\hat{O}_{S_1},\hat{O}_{S_2}]\neq\hat{0}$ would become usable in logical inference. Even though the commutator $[\hat{O}_{S_1},\hat{O}_{S_2}]$ has an unambiguous interpretation with respect to the mathematical formalism of quantum mechanics, it cannot be formalized into a statement with a precise interpretation of purely logical character determined from logic alone. Simply put, the commutator $[\hat{O}_{S_1},\hat{O}_{S_2}]$ cannot be translated into a language of logic.\\

\noindent But as the commutator $[\hat{O}_{S_1},\hat{O}_{S_2}]$ does not belong to a formal system of logic (that is used for constructing logical sentences from axioms according to logical calculus), inserting it in the truth functions is equivalent to adding an extra postulate to the instrumentalist description of quantum theory and – in turn – offering a new interpretation of quantum mechanics (so-called \emph{quantum-logic interpretation}).\\

\noindent To get rid of the commutator $[\hat{O}_{S_1},\hat{O}_{S_2}]$ in logical calculus (and so an extra postulate it brings), one may drop any possibility to determine a truth value of a compound statement by truth values of its constituents; explicitly, one may declare\smallskip

\begin{equation}  %{Eq.28}
   [\mkern-3.3mu[
      S_1{\ast}S_2
   ]\mkern-3.3mu]
   \mkern-2mu
   \neq
   f_{\mkern-2mu\ast}
   \mkern-2mu
   \left(
      [\mkern-3.3mu[S_1]\mkern-3.3mu]
      ,
      [\mkern-3.3mu[S_2]\mkern-3.3mu]
   \right)
   \;\;\;\;  .
\end{equation}
\smallskip

\noindent This is the same as to assert that a non-classical propositional logic, in which compound statements are not truth functions (similar to a non-distributive logic or the calculus of partial propositional functions), is the “true” logic.\\

\noindent On the other hand, given that the ``true'' logic is not classical, how does one explain that on some occasions, the logical connectives $\sqcup$ and $\sqcap$ distribute one over the other and are truth-functional?\\

\noindent What is more, it makes perfect sense to assume that if a physical system undergoes a process that causes its behavior to become classical, then the mathematical model of reasoning used to appraise statements regarding this system must undergo a transition from a non-distributive propositional logic to classical, i.e., distributive, logic. So, the question is, how can one explain such a transition?\\

\noindent The difficulty in this explanation is that the classical limit $\hbar{\mkern1.5mu\to\mkern1.5mu}0$ wherein any two observables $\hat{O}_{S_1}$ and $\hat{O}_{S_2}$ may be regarded as commutative (i.e., classical) cannot be translated into a language of logic. As a result, bringing this limit into logical calculus would once again be equivalent to adding an extra postulate to the instrumentalist description of quantum mechanics.\\

\noindent So, taking the approach to modify truth-functionality of the logical operators $\sqcup$ and $\sqcap$, one has no choice but to add an extra postulate to the instrumentalist description of quantum theory justifying either the presence of the commutator $[\hat{O}_{S_1},\hat{O}_{S_2}]$ in logical calculus or its absence therein!\\

\section{Proposal to deny bivalence of statements}  %{<-------------------------------------------------------------------------------------------------Section V}

\noindent Modifying (\ref{TV}), namely, allowing a statement to have a truth value other than “true” or “false”, can also nullify negative construal of quantum superposition.\\

\noindent To make this evident, assume that before observation, both the statement $H$ and the statement $T$ have the third truth value denoted U. This value can be interpreted as a real number lying between 0 and 1, say, $\rfrac{1}{2}$ (in agreement with {\L}ukasiewicz logic $\mathrm{L}_3$ \cite{Lukasiewicz}). The nature of the third truth value is supposed to be the same as the one of the truth values T and F (if not, it would be hard to guarantee the consistency and unambiguity of what constitutes a set of the truth values \cite{Gottwald}). More formally, this assumption can be written as\smallskip

\begin{equation}  %{Eq.29}
   t
   \in
   [t_r,t_o)
   \mkern-3.3mu
   :
   \mkern4mu
   \begingroup
   \begin{array}{r c c c}
      v:
      &
      \mathbb{S}
      &
      \to
      &
      \mathbb{B}_3
      \\[5pt]
      \hfill
      &
      H
      &
      \mapsto
      &
      \rfrac{1}{2}
      \\[5pt]
      \hfill
      &
      T
      &
      \mapsto
      &
      \rfrac{1}{2}
   \end{array}
   \endgroup
   \;\;\;\;  ,
\end{equation}
\smallskip

\noindent where the first part is read “$v$ is a total function from $\mathbb{S}$ to $\mathbb{B}_3=\{0,\rfrac{1}{2},1\}$”, while the second and the third parts declare that both $H$ and $T$, the elements of $\mathbb{S}$, map onto $\rfrac{1}{2}$, the element of $\mathbb{B}_3$.\\

\noindent Introducing at this step {\L}ukasiewicz conjunction and disjunction \cite{Pykacz17}, which interpret the logical connectives $\sqcup$ and $\sqcap$ joining two statements $S_1$ and $S_2$ in the following way\smallskip

\begin{equation}  %{Eq.30}
   [\mkern-3.3mu[
      S_1
      {\mkern1.5mu\sqcap\mkern1.5mu}
      S_2
   ]\mkern-3.3mu]         
   =
   \max
   \mkern-2mu
   \left(
      [\mkern-3.3mu[S_1]\mkern-3.3mu]
      +
      [\mkern-3.3mu[S_2]\mkern-3.3mu]
      -
      1
      ,
      0
   \right)
   \;\;\;\;  ,
\end{equation}
\\[-36pt]

\begin{equation}  %{Eq.31}
   [\mkern-3.3mu[
      S_1
      {\mkern1.5mu\sqcup\mkern1.5mu}
      S_2
   ]\mkern-3.3mu]         
   =
   \min
   \mkern-2mu
   \left(
      [\mkern-3.3mu[S_1]\mkern-3.3mu]
      +
      [\mkern-3.3mu[S_2]\mkern-3.3mu]
      ,
      1
   \right)
   \;\;\;\;  ,
\end{equation}
\smallskip

\noindent one finds that\smallskip

\begin{equation}  %{Eq.32}
   t
   \in
   [t_r,t_o)
   \mkern-3.3mu
   :
   \mkern12mu
   [\mkern-3.3mu[
      H
      {\mkern1.5mu\sqcap\mkern1.5mu}
      T
   ]\mkern-3.3mu]         
   =
   \max
   \mkern-2mu
   \left(
      \rfrac{1}{2}
      +
     \rfrac{1}{2}
      -
      1
      ,
      0
   \right)
   =
   0
   \;\;\;\;  .
\end{equation}
\smallskip

\noindent Furthermore, on condition that after observation the statements $H$ and $T$ obey the principle of bivalence and are not able to be true together, one gets\smallskip

\begin{equation}  %{Eq.33}
   t
   \in
   [t_o,+\infty)
   \mkern-3.3mu
   :
   \mkern12mu
   [\mkern-3.3mu[
      H
      {\mkern1.5mu\sqcap\mkern1.5mu}
      T
   ]\mkern-3.3mu]         
   =
   \max
   \mkern-2mu
   \left(
      [\mkern-3.3mu[H]\mkern-3.3mu]
      +
      [\mkern-3.3mu[T]\mkern-3.3mu]
      -
      1
      ,
      0
   \right)
   =
   0
   \;\;\;\;  .
\end{equation}
\smallskip

\noindent In this way, the compound statement $H{\mkern1.5mu\sqcap\mkern1.5mu}T$ has one and the same truth value, explicitly, “false”, over the whole interval $[t_r,+\infty)$, which causes “quantum fundamentalism” and “macroscopic realism” to be consistent and the problem of definite outcomes to disappear.\\

\noindent But another problem arises instead: Why do the statements $H$ and $T$ obey the principle of bivalence after observation? Put differently, if the third truth value $\rfrac{1}{2}$ has the same nature as the binary truth values 1 and 0 have, why does only it get destroyed once the coin is observed?\\

\noindent One may suggest that the truth value $\rfrac{1}{2}$ does not survive observation because of the existence of hidden variables, or interaction of human mind with the surrounding physical world, or something of that kind. But whatever the case, any of such suggestions would inevitably imply bringing a new postulate (or postulates) into the instrumentalist description of quantum theory.\\

\section{Proposal to modify bivalent predicate logic}  %{<-------------------------------------------------------------------------------------------------Section VI}

\noindent Lastly, to render void negative construal of quantum superposition, one can modify both expressions (\ref{TV}) and (\ref{BCON}).\\

\noindent As an example of such a twofold modification, let us review \emph{the proposal of multivalued predicate logic} presented in \cite{Pykacz15}. In accordance with this proposal, a meaningful declarative expression regarding a physical system, such as \text{\guillemotleft}The categorical property of the coin is “heads up”\text{\guillemotright}, is not a statement per se but rather a predicate or a propositional (statemental) function, which cannot have a truth value of its own. Only when a state of a system is assigned to such an expression, it may become a statement capable of being true or false.\\

\noindent To make the aforesaid functionality explicit, let us put a direct mention of a state into each declarative expression regarding a physical system. Consider, for example, the following statemental function \text{\guillemotleft}If the state of the coin is “state”, then the categorical property of the coin is “heads up”\text{\guillemotright} denoted by $H(\text{``state''})$. The wording of this function can be abbreviated to \text{\guillemotleft}If “state” is true, then “heads up” is true\text{\guillemotright}. In the same way, $T(\text{``state''})$ denotes the statemental function \text{\guillemotleft}If “state” is true, then “tails up” is true\text{\guillemotright}. When “state” is given, say, “heads up”, these functions turn into the statements \text{\guillemotleft}If “heads up” is true, then “heads up” is true\text{\guillemotright} and \text{\guillemotleft}If “heads up” is true, then “tails up” is true\text{\guillemotright} whose truth values are $[\mkern-3.3mu[H(\text{``heads up''})]\mkern-3.3mu]=1$ and $[\mkern-3.3mu[T(\text{``heads up''})]\mkern-3.3mu]=0$.\\

\noindent It is important to remember that, despite the presence of the connective “if … then” in the above statements, they are atomic. Really, had the opposite been true, an assignment of truth values to expressions like \text{\guillemotleft}“heads up” is true\text{\guillemotright} and \text{\guillemotleft}“tails up” is true\text{\guillemotright} would have been possible without knowing the state of the coin.\\

\enlargethispage{\baselineskip}
\enlargethispage{\baselineskip}
\noindent Let us introduce $\Pr[H(\text{``state''})]$, the probability that the experiment, which is designed to check whether the upper side of the coin in the state denoted by “state” is “heads”, will give an affirmative answer. In line with the instrumentalist description of quantum theory, the probability such as this is equal to the “matrix element” $\langle\Psi|\hat{O}_{H}|\Psi\rangle$, where $|\Psi\rangle$ is a vector of the Hilbert space that describes the state of the coin “state”, while $\hat{O}_{H}$ is the observable that encodes $H(\text{``state''})$ when “state” is given. Accordingly, one can put down the following:\smallskip

\begin{equation} \label{QPRH} %{Eq.34}
   \Pr\mkern-1.5mu\left[H(\text{``state''})\right]
   =
   \langle\Psi|\hat{O}_{H}|\Psi\rangle
   \;\;\;\;  .
\end{equation}
\smallskip

\noindent As postulated in \cite{Pykacz15}, the truth value of the statemental function $H$ on the given “state” must be the same as the above probability, that is,\smallskip

\begin{equation} \label{PUCK1} %{Eq.35}
   [\mkern-3.3mu[H(\text{``state''})]\mkern-3.3mu]
   =
   \langle\Psi|\hat{O}_{H}|\Psi\rangle
   \;\;\;\;  .
\end{equation}
\smallskip

\noindent By the same token,\smallskip

\begin{equation} \label{QPRT} %{Eq.36}
   \Pr\mkern-1.5mu\left[T(\text{``state''})\right]
   =
   \langle\Psi|\hat{O}_{T}|\Psi\rangle
   \;\;\;\;  ,
\end{equation}
\smallskip

\noindent and so\smallskip

\begin{equation} \label{PUCK2} %{Eq.37}
   [\mkern-3.3mu[T(\text{``state''})]\mkern-3.3mu]
   =
   \langle\Psi|\hat{O}_{T}|\Psi\rangle
   \;\;\;\;  .
\end{equation}
\smallskip

\noindent For the duration of the period $[t_r,t_o)$, the coin is in the superposed state “heads up and tails up” described by the vector $|\Psi_{R}\rangle=c_{H}|\Psi_{H}\rangle+c_{T}|\Psi_{T}\rangle$, where $|c_{H}|^2+|c_{T}|^2=1$. As long as $\langle\Psi_{T}|\Psi_{H}\rangle=0$, the “matrix element” $\langle\Psi_{R}|\hat{O}_{H}|\Psi_{R}\rangle$ is\smallskip

\begin{equation}  %{Eq.38}
   \Big(
      c_{H}^{\ast}\langle\Psi_{H}|
      +
      c_{T}^{\ast}\langle\Psi_{T}|
   \Big)
   \mkern2mu
   \hat{O}_{H}
   \mkern-0.5mu
   \Big(
      c_{H}|\Psi_{H}\rangle
      +
      c_{T}|\Psi_{T}\rangle
   \Big)
   =
   \Big(
      c_{H}^{\ast}\langle\Psi_{H}|
      +
      c_{T}^{\ast}\langle\Psi_{T}|
   \Big)
   \mkern2mu
   c_{H}|\Psi_{H}\rangle
   =
   |c_{H}|^2
   \;\;\;\;  .
\end{equation}
\smallskip

\noindent Similarly, the “matrix element” $\langle\Psi_{R}|\hat{O}_{T}|\Psi_{R}\rangle$ is equal to $|c_{T}|^2$. Therefore, when “state” in the statemental functions $H$ and $T$ is substituted for “heads up and tails up”, these functions – unlike predicates in bivalent predicate logic – can return more than two truth values (that is why the approach discussed here is called the proposal of multivalued predicate logic):\smallskip

\begin{equation}  %{Eq.39}
   [\mkern-3.3mu[H(\text{``heads up and tails up''})]\mkern-3.3mu]
   =
   |c_{H}|^2
   \;\;\;\;  ,
\end{equation}
\\[-36pt]

\begin{equation}  %{Eq.40}
   [\mkern-3.3mu[T(\text{``heads up and tails up''})]\mkern-3.3mu]
   =
   |c_{T}|^2
   \;\;\;\;  .
\end{equation}
\smallskip

\noindent As a final point, after using the {\L}ukasiewicz truth functions defined as follows\smallskip

\begin{equation}  %{Eq.41}
   [\mkern-3.3mu[
      S_1{\sqcap}S_2(\text{``state''})
   ]\mkern-3.3mu]
   \mkern-2mu
   =
   \left\{
      \begingroup\SmallColSep
      \begin{array}{l l}
         \max
         \mkern-2mu
         \Big(
            [\mkern-3.3mu[S_1(\text{``state''})]\mkern-3.3mu]
            +
            [\mkern-3.3mu[S_2(\text{``state''})]\mkern-3.3mu]
            -
            1
            ,
            0
         \Big)
         &
         \mkern3mu
         ,
         \mkern12mu
         \left[
            \hat{O}_{S_1}
            ,
            \hat{O}_{S_2}
         \right]
         \mkern-2mu
         =
         \hat{0}
         \\[5pt]
         \text{undefined}
         &
         \mkern3mu
         ,
         \mkern12mu
         \left[
            \hat{O}_{S_1}
            ,
            \hat{O}_{S_2}
         \right]
         \mkern-2mu
         \neq
         \hat{0}
      \end{array}
      \endgroup   
   \right.   
   \;\;\;\;  ,
\end{equation}
\\[-20pt]

\begin{equation}  %{Eq.42}
   [\mkern-3.3mu[
      S_1{\sqcup}S_2(\text{``state''})
   ]\mkern-3.3mu]
   \mkern-2mu
   =
   \left\{
      \begingroup\SmallColSep
      \begin{array}{l l}
         \min
         \mkern-2mu
         \Big(
            [\mkern-3.3mu[S_1(\text{``state''})]\mkern-3.3mu]
            +
            [\mkern-3.3mu[S_2(\text{``state''})]\mkern-3.3mu]
            ,
            1
         \Big)
         &
         \mkern3mu
         ,
         \mkern12mu
         \left[
            \hat{O}_{S_1}
            ,
            \hat{O}_{S_2}
         \right]
         \mkern-2mu
         =
         \hat{0}
         \\[5pt]
         \text{undefined}
         &
         \mkern3mu
         ,
         \mkern12mu
         \left[
            \hat{O}_{S_1}
            ,
            \hat{O}_{S_2}
         \right]
         \mkern-2mu
         \neq
         \hat{0}
      \end{array}
      \endgroup   
   \right.   
   \;\;\;\;  ,
\end{equation}
\smallskip

\noindent where $S_1{\mkern2mu\ast\mkern2mu}S_2(\text{``state''})$ is used to shorten the combination $S_1(\text{``state''}){\mkern2mu\ast\mkern2mu}S_2(\text{``state''})$, one finds that negative construal of quantum superposition is invalid:\smallskip

\begin{equation}  %{Eq.43}
   [\mkern-3.3mu[
      H{\mkern1.5mu\sqcap\mkern1.5mu}T
      (\text{``heads up and tails up''})
   ]\mkern-3.3mu]
   =
   \max
   \mkern-2.5mu
   \left(
      |c_{H}|^2
      +
      |c_{T}|^2
      -
      1
      ,
      0
   \right)
   =
   0
   \;\;\;\;  .
\end{equation}
\smallskip

\noindent Be that as it may, one must recognize that the last result implies that the axioms of Kolmogorov’s probability theory are wrong.\\

\noindent To reveal this, observe that the equalities like (\ref{QPRH}) and (\ref{PUCK1}), as well as (\ref{QPRT}) and (\ref{PUCK2}), are collateral to the generic postulate\smallskip

\begin{equation} \label{GAS} %{Eq.44}
   \Pr\mkern-1.5mu\left[S(\text{``state''})\right]
   =
   [\mkern-3.3mu[S(\text{``state''})]\mkern-3.3mu]
   \;\;\;\;  ,
\end{equation}
\smallskip

\noindent where $S(\text{``state''})$ denotes an arbitrary statemental function. Using this postulate to substitute truth values for probabilities in the {\L}ukasiewicz truth functions, one gets\smallskip

\begin{equation}  %{Eq.45}
   \Pr\mkern-1.5mu\Big[
      S_1{\mkern1.5mu\sqcap\mkern1.5mu}S_2
      (\text{``state''})
   \Big]
   =
   \max
   \mkern-2.5mu
   \Big(
      \Pr\mkern-1.5mu\left[S_1(\text{``state''})\right]
      +
      \Pr\mkern-1.5mu\left[S_2(\text{``state''})\right]
      -
      1
      ,
      0
   \Big)
   \;\;\;\;  .
\end{equation}
\smallskip

\noindent However, the above contradicts the Kolmogorov axioms, which maintain that the probability of $S_1{\mkern1.5mu\sqcap\mkern1.5mu}S_2(\text{``state''})$ cannot be expressed as a function of $\Pr[S_1(\text{``state''})]$ and $\Pr[S_2(\text{``state''})]$.\\

\noindent To sum up: In order to cancel negative construal of quantum superposition by way of the proposal of multivalued predicate logic, the postulate of equivalence between truth values and probability values must be added to the instrumentalist description of quantum theory, even though such an equivalence is tantamount to discarding one of the common interpretations of probability!\\

\section{Logic is not empirical}  %{<-------------------------------------------------------------------------------------------------Section VII}

\noindent A generalization that one might infer from the previous sections is that no rejection or modification of rules of logical systems (such as propositional and predicate logics) is capable of canceling negative construal of quantum superposition without adding an extraneous postulate (or postulates) to the instrumentalist description of quantum theory. The reason is that \emph{at least some statemental functions in this description are not analytic}.\\

\noindent To clarify this observation, let us recall that a predicate $S$ on a subject $X$ is analytic if the substitution of any entity for $X$ (from the set of values $X$ may take), say $x$, will produce an analytic statement $S(x)$, i.e., a statement whose truth or falsity can be known merely by knowing the meaning of its constituent terms \cite{Rey}. One may say that $S(X)$ is analytic if no fact is needed to accept that $S(x)$ is true or false.\\

\noindent Let us take as an example the statements \text{\guillemotleft}If “heads up” is true, then “heads up” is true\text{\guillemotright} and \text{\guillemotleft}If “heads up” is true, then “tails up” is true\text{\guillemotright} mentioned before. One need not consult experience to determine that of these two, the former is true. Furthermore, after clarifying that “tails up” is \emph{not “heads up”}, the latter becomes \text{\guillemotleft}If “heads up” is true, then \emph{not “heads up”} is true\text{\guillemotright}, which is false solely by virtue of the meaning of the adverb “not”. Hence, both statements can be classified as analytic.\\

\noindent By contrast, the statement \text{\guillemotleft}If ”heads up and tails up” is true, then “heads up” is true\text{\guillemotright} is not analytic. This is because the instrumentalist description of quantum theory does not assign a meaning to superposition of states. Especially, this description does not define the meaning of the coordinating conjunction “and” in the phrase “heads up and tails up”. As a result, within the description such as this, the aforesaid statement cannot be analytically true or analytically false, that is, true or false just in virtue of knowing what each word in the statement means.\\

\noindent Since for any physical system the most general state is a superposition of all (theoretically possible) states, one can conclude that in the quantum instrumentalist description, statemental functions like \text{\guillemotleft}If the state of a physical system is “state”, then the system’s categorical property is “$s$”$\mkern0.5mu$\text{\guillemotright}, denoted by $S(\text{``state''})$, cannot generally be analytic.\\

\noindent Oppositely, such functions are analytic in the mathematical formalism of classical mechanics. In fact, the concept of categorical property therein is synonymous with a collection of states of a physical system that have some property in common. Once the said collection is substituted for its synonymic term, experience is no longer necessary to decide $[\mkern-3.3mu[S(\text{``state''})]\mkern-3.3mu]$: Any given state of the system will or will not be an element of the specified collection merely on account of the definition of the collection.\\

\noindent This suggests that by substitution of synonyms, classical mechanics could be all derived from set theory and mapping, and therefore from logic (providing mathematics is an extension of logic). Accordingly, logic can be provable or verifiable by classical mechanical experience.\\

\noindent However, the same does not apply to quantum experience. As it is not the case that all statemental functions in the quantum instrumentalist description are analytic, this description cannot be reducible to logic. On condition that the said description need not be supplemented with additional postulates to explain the physical universe, one can maintain that the epistemological status of logical predicates – analytic by their nature – differs from that of statemental functions about the physical universe, not analytic in general. To put it in a nutshell, if “quantum fundamentalism” holds true, logic cannot be considered empirical.\\

\noindent Still, one might try to introduce an additional postulate, which would define the meaning of “and” in superposition of states. In that case, the quantum mathematical formalism – amended by this additional postulate – might be capable of being reduced to some type of non-classical logic.\\

\noindent Nonetheless, the point to bear in mind is that the resultant quantum-logic model would represent nothing more than an interpretation of quantum mechanics and would not act in the place of its bare instrumentalist description.\\

\section{The alternative model of reasoning}  %{<-------------------------------------------------------------------------------------------------Section VIII}

\noindent Since logic is not empirical and so not capable of canceling negative construal of quantum superposition on its own, one may seek to employ an alternative model of reasoning to make “quantum fundamentalism” and “macroscopic realism” nonconflicting without adding a new hypothesis to the instrumentalist description of quantum theory.\\

\noindent Recall that in its simplest form, propositional logic can be considered as the study of “truth tables”, in which one or more outputs (compound statements) depend on a combination of connectives (“and”, “or”, “not”, etc.) and the input values (“atoms”) \cite{Enderton}. Therewithal, propositional calculus is based on the implicit assumption that \emph{an interpretation of all “atoms”} – i.e., assigning a truth value to each of them – \emph{can be done in a completely free manner}, to be exact, without being controlled or restricted by laws of the world and conditional solely upon one's discretion.\\

\noindent It is also true for intuitionistic logic wherein, instead of assigning each ``atom'' a value of a Boolean algebra, values from a Heyting algebra are used \cite{Moschovakis}. To demonstrate this point, let us consider a Heyting algebra whose elements are open subsets $(x^{\prime},x^{\prime\prime})$ of the real line $\mathbb{R}$. In such an algebra, the valuation for an ``atom'' $S$ takes the form\smallskip

\begin{equation}  %{Eq.46}
   [\mkern-3.3mu[S]\mkern-3.3mu]
   =
   \left(x^{\prime},x^{\prime\prime}\right)
   =
   \Big\{
      x
      \in
      \mathbb{R}
      \Big|
         x^{\prime}
         <
         x
         <
         x^{\prime\prime}
      \Big.
   \Big\}
   \;\;\;\;  .
\end{equation}
\smallskip

\noindent Because the endpoints $x^{\prime}$ and $x^{\prime\prime}$ are arbitrary, the truth value of $S$ is decided freely between $(x^{\prime},x^{\prime})=\emptyset=[\mkern-3.3mu[\bot]\mkern-3.3mu]$ and $(-\infty,+\infty)=\mathbb{R}=[\mkern-3.3mu[\top]\mkern-3.3mu]$. And in case $x^{\prime}$ and $x^{\prime\prime}$ stay undetermined, the ``atom'' $S$ remains of unknown truth value. Accordingly, an interpretation of all the ``atoms'' in intuitionistic logic can be free as much as it does in propositional logic.\\

\noindent Given the above, any model of reasoning, which did not accept the assumption of a free interpretation of the ``atoms'' as valid, would be an alternative to logic. Specifically, this might be a mathematical model which requires \emph{proof to assign either truth value to each “atom”}. In this model, if there is no proof for some “atom”, it will exist \emph{without a truth value at all}.\\

\noindent Importantly, the required proof is not allowed to be contingent (had it been so, the truth values “true” and “false” would not have been objective). Hence, the proof must be necessary, i.e., such that it cannot be denied without contradiction. Equating \emph{necessary} with \emph{analytic} \cite{Sloman}, the last implies that what counts as proof in the alternative model of reasoning must be analytic.\\

\noindent On the other hand, a mathematical truth – i.e., an atomic truthful statement of a mathematical relation (such as equality or set membership) between meaningful expressions (to wit: symbols or combinations of symbols representing a value, a function, an object, or the like) – is analytic since it is determined solely by the conventions of mathematical axioms and not contingent on observation of facts. It follows then that in the role of the analytic proof, which compels acceptance of the truthfulness of the “atom”, the alternative model of reasoning takes a mathematical truth, synonymous with the “atom”. Symmetrically, a mathematical truth, antonymous to the “atom”, is taken as the analytic proof of its falsity.\\

\noindent By the same token, the truth of mathematical predicates, synonymous and antonymous with the atomic statemental function (i.e., the one which becomes the ``atom'' when function’s variables are given definite admissible values), is taken as the analytic proof of the truthfulness or falsity of the function. So, if neither of these proofs is present, that is, if the mathematical synonyms and antonyms are all false or nonexistent, then the atomic function is regarded as having no truth value.\\

\noindent To illustrate how the alternative model of reasoning works, let’s go again to the atomic statemental function $H(\text{``state''})$ worded as \text{\guillemotleft}If ``state'' is true, then “heads up” is true\text{\guillemotright}. In the instrumentalist description of quantum theory, the categorical properties “heads up” and “tails up” are respectively represented by the rays $\mathcal{E}_{H}$ and $\mathcal{E}_{T}$ (i.e., 1-dimensional vector subspaces of $\mathcal{H}$), which consist of vectors related to the (nonzero) vectors $|\Psi_{H}\rangle$ and $|\Psi_{T}\rangle$ by a complex multiplicative factor, say $c$, that is, $\mathcal{E}_{H}=\{c|\Psi_{H}\rangle\}$ and $\mathcal{E}_{T}=\{c|\Psi_{T}\rangle\}$. Hence, the mathematical predicates $\eta(|\Psi\rangle)$ and $\tau(|\Psi\rangle)$, specifically,\smallskip

\begin{equation}  %{Eq.47}
   \eta(|\Psi\rangle)
   =
   \Big\{
      |\Psi\rangle
      \in
      \mathcal{H}
      \Big|
         |\Psi\rangle
         \in
         \mathcal{E}_{H}
      \Big.
   \Big\}
   \;\;\;\;  ,
\end{equation}
\\[-36pt]

\begin{equation}  %{Eq.48}
   \tau(|\Psi\rangle)
   =
   \Big\{
      |\Psi\rangle
      \in
      \mathcal{H}
      \Big|
         |\Psi\rangle
         \in
         \mathcal{E}_{T}
      \Big.
   \Big\}
   \;\;\;\;  ,
\end{equation}
\smallskip

\noindent are the synonym and antonym, in that order, for the function $H(\text{``state''})$. Each of these predicates is true if a vector $|\Psi\rangle$ representing “state” belongs to the set being defined and false otherwise.\\

\noindent Being the opposite of one to the other, the predicates $\eta(|\Psi\rangle)$ and $\tau(|\Psi\rangle)$ cannot be true together, in symbols:\smallskip

\begin{equation}  %{Eq.49}
   |\Psi\rangle
   \neq
   0
   \mkern-3.3mu
   :
   \mkern12mu
   [\mkern-3.3mu[
      \eta(|\Psi\rangle)
      {\mkern1.5mu\sqcap\mkern1.5mu}
      \tau(|\Psi\rangle)
   ]\mkern-3.3mu]         
   =
   \min
   \mkern-2mu
   \left(
      [\mkern-3.3mu[\eta(|\Psi\rangle)]\mkern-3.3mu]
      ,
      [\mkern-3.3mu[\tau(|\Psi\rangle)]\mkern-3.3mu]
   \right)
   =
   0
   \;\;\;\;  .
\end{equation}
\smallskip

\noindent Consequently, the atomic function $H(\text{``state''})$ becomes true in each situation wherein $\eta(|\Psi\rangle)$ is true. Contrastively, the said function gets false whenever $\tau(|\Psi\rangle)$ is true. In case $\eta(|\Psi\rangle)$ and $\tau(|\Psi\rangle)$ are both false, $H(\text{``state''})$ remains without value. Symbolically, this can be shown as follows:\smallskip

\begin{equation} \label{SV} %{Eq.50}
   [\mkern-3.3mu[H(\text{``state''})]\mkern-3.3mu]
   =
   \left\{
      \begingroup\SmallColSep
      \begin{array}{r l}
         1
         &
         \mkern3mu
         ,
         \mkern12mu
         [\mkern-3.3mu[\eta(|\Psi\rangle)]\mkern-3.3mu]
         =
         1
         \\[3pt]
         0
         &
         \mkern3mu
         ,
         \mkern12mu
         [\mkern-3.3mu[\tau(|\Psi\rangle)]\mkern-3.3mu]
         =
         1
         \\[3pt]
         \textnormal{NULL}
         &
         \mkern3mu
         ,
         \mkern12mu
         [\mkern-3.3mu[\eta(|\Psi\rangle)]\mkern-3.3mu]
         =
         [\mkern-3.3mu[\tau(|\Psi\rangle)]\mkern-3.3mu]
         =
         0
      \end{array}
      \endgroup   
   \right.
   \;\;\;\;  ,
\end{equation}
\smallskip

\noindent where NULL is the symbol that is used to denote “nothingness”.\\

\noindent For example, let “state” be “heads up” represented by $|\mkern-2mu\Psi_{\mkern-2mu{H}}\mkern-2mu\rangle$. Then, as stated above, $[\mkern-3.3mu[H(\text{``heads up''})]\mkern-3.3mu]=1$ because $[\mkern-3.3mu[\eta(|\Psi_{\mkern-2mu{H}}\rangle)]\mkern-3.3mu]=1$. But if “state” is “tails up”, one finds $[\mkern-3.3mu[H(\text{``tails up''})]\mkern-3.3mu]=0$ because $[\mkern-3.3mu[\tau(|\Psi_{\mkern-2mu{T}}\rangle)]\mkern-3.3mu]=1$.\\

\noindent Now, suppose that “state” is “heads up and tails up”. This state is represented by the vector $|\mkern-0.5mu\Psi_{\mkern-0.5mu{R}}\mkern-0.5mu\rangle=c_{H} |\mkern-0.5mu\Psi_{\mkern-0.5mu{H}}\mkern-0.5mu\rangle+c_{T}|\mkern-0.5mu\Psi_{\mkern-0.5mu{T}}\mkern-0.5mu\rangle$ on which both predicates $\eta$ and $\tau$ are false: $[\mkern-3.3mu[\eta(|\Psi_{\mkern-0.5mu{R}}\rangle)]\mkern-3.3mu]=[\mkern-3.3mu[\tau(|\Psi_{\mkern-0.5mu{R}}\rangle)]\mkern-3.3mu]=0$. Accordingly, $H(\text{``heads up and tails up''})$ has no truth value: $[\mkern-3.3mu[H(\text{``heads up and tails up''})]\mkern-3.3mu]=\textnormal{NULL}$.\\

\noindent Likewise, the atomic statemental function $T(\text{``state''})$, whose valuation is defined by\smallskip

\begin{equation}  %{Eq.51}
   [\mkern-3.3mu[T(\text{``state''})]\mkern-3.3mu]
   =
   \left\{
      \begingroup\SmallColSep
      \begin{array}{r l}
         1
         &
         \mkern3mu
         ,
         \mkern12mu
         [\mkern-3.3mu[\tau(|\Psi\rangle)]\mkern-3.3mu]
         =
         1
         \\[3pt]
         0
         &
         \mkern3mu
         ,
         \mkern12mu
         [\mkern-3.3mu[\eta(|\Psi\rangle)]\mkern-3.3mu]
         =
         1
         \\[3pt]
         \textnormal{NULL}
         &
         \mkern3mu
         ,
         \mkern12mu
         [\mkern-3.3mu[\tau(|\Psi\rangle)]\mkern-3.3mu]
         =
         [\mkern-3.3mu[\eta(|\Psi\rangle)]\mkern-3.3mu]
         =
         0
      \end{array}
      \endgroup   
   \right.
   \;\;\;\;  ,
\end{equation}
\smallskip

\noindent returns no value if “state” is “heads up and tails up”.\\

\noindent Let the alternative model of reasoning be closed under connectives. This implies that for any two statements $S_1$ and $S_2$, the compound $S_1{\ast}S_2$ is also a statement. Therefore, mathematical truths synonymous and antonymous with $S_1{\ast}S_2$ can be taken as the analytic proof of the truthfulness or falsity of the said compound.\\

\noindent Consider as an example the compound categorical property “heads up” \emph{and} “tails up”. To denote the subspace of $\mathcal{H}$, which represents this property, the symbol $\mathcal{E}_{H{\mkern1.5mu\sqcap\mkern1.5mu}T}$ will be used. According to the instrumentalist description of quantum theory, the subspace $\mathcal{E}_{H{\mkern1.5mu\sqcap\mkern1.5mu}T}$ is the intersection of two rays $\mathcal{E}_{H}$ and $\mathcal{E}_{T}$, denoted by $\mathcal{E}_{H}{\mkern1mu\cap\mkern2mu}\mathcal{E}_{T}$. In symbols,\smallskip

\begin{equation}  %{Eq.52}
   \mathcal{E}_{H{\mkern1.5mu\sqcap\mkern1.5mu}T}
   =
   \mathcal{E}_{H}
   \cap
   \mathcal{E}_{T}
   =
   \{0\}
   \;\;\;\;  .
\end{equation}
\smallskip

\noindent As to the compound categorical property “heads up” \emph{or} “tails up”, it is represented by the subspace of $\mathcal{H}$ denoted by $\mathcal{E}_{H{\mkern1.5mu\sqcup\mkern1.5mu}T}$, which is the span of the union of the rays $\mathcal{E}_{H}$ and $\mathcal{E}_{T}$, according to the quantum instrumentalist description. Consequently,\smallskip

\begin{equation}  %{Eq.53}
   \mathcal{E}_{H{\mkern1.5mu\sqcup\mkern1.5mu}T}
   =
   \mathcal{E}_{H}
   +
   \mathcal{E}_{T}
   =
   \mathcal{H}
   \;\;\;\;  ,
\end{equation}
\smallskip

\noindent see \cite{Halmos, Piron} detailing the above equivalences. So, the predicates\smallskip

\begin{equation}  %{Eq.54}
   \sigma_{H{\mkern1.5mu\sqcap\mkern1.5mu}T}(|\Psi\rangle)
   =
   \Big\{
      |\Psi\rangle
      \in
      \mathcal{H}
      \Big|
         |\Psi\rangle
         \in
         \{0\}
      \Big.
   \Big\}
   \;\;\;\;  ,
\end{equation}
\\[-36pt]

\begin{equation}  %{Eq.55}
   \alpha_{H{\mkern1.5mu\sqcap\mkern1.5mu}T}(|\Psi\rangle)
   =
   \Big\{
      |\Psi\rangle
      \in
      \mathcal{H}
      \Big|
         |\Psi\rangle
         \in
         \mathcal{H}
      \Big.
   \Big\}
   \;\;\;\;  ,
\end{equation}
\smallskip

\noindent are the synonym and antonym, in that order, for the function \text{\guillemotleft}If ``state'' is true, then “heads up” \emph{and} “tails up” is true\text{\guillemotright}, denoted $H{\mkern1.5mu\sqcap\mkern1.5mu}T(\text{``state''})$, and, in reverse order, for the function \text{\guillemotleft}If ``state'' is true, then “heads up” \emph{or} “tails up” is true\text{\guillemotright} denoted $H{\mkern1.5mu\sqcup\mkern1.5mu}T(\text{``state''})$:\smallskip

\begin{equation}  %{Eq.56}
   \sigma_{H{\mkern1.5mu\sqcup\mkern1.5mu}T}(|\Psi\rangle)
   =
   \alpha_{H{\mkern1.5mu\sqcap\mkern1.5mu}T}(|\Psi\rangle)
   \;\;\;\;  ,
\end{equation}
\\[-37pt]

\begin{equation}  %{Eq.57}
   \alpha_{H{\mkern1.5mu\sqcup\mkern1.5mu}T}(|\Psi\rangle)
   =
   \sigma_{H{\mkern1.5mu\sqcap\mkern1.5mu}T}(|\Psi\rangle)
   \;\;\;\;  .
\end{equation}
\smallskip

\noindent For all vectors $|\Psi\rangle$ of a Hilbert space, $[\mkern-3.3mu[\alpha_{H{\mkern1.5mu\sqcap\mkern1.5mu}T}(|\Psi\rangle)]\mkern-3.3mu]\neq{0}$. Consequently, the predicates $\sigma_{H{\mkern1.5mu\sqcap\mkern1.5mu}T}(|\Psi\rangle)$ and $\alpha_{H{\mkern1.5mu\sqcap\mkern1.5mu}T}(|\Psi\rangle)$ are not false together. Furthermore, providing all vectors describing states of a physical system are nonzero, $\sigma_{H{\mkern1.5mu\sqcap\mkern1.5mu}T}(|\Psi\rangle)$ is the empty set while $\alpha_{H{\mkern1.5mu\sqcap\mkern1.5mu}T}(|\Psi\rangle)$ is the universal set. Thus, using the symbolic expression $\sqcap\mkern-7.5mu\sqcup\in\{\sqcap,\sqcup\}$, the valuation of $H{\mkern1.5mu\sqcap\mkern1.5mu}T(\text{``state''})$ and $H{\mkern1.5mu\sqcup\mkern1.5mu}T(\text{``state''})$ can be presented as follows:\smallskip

\begin{equation}  %{Eq.58}
   [\mkern-3.3mu[
      H{\mkern1.5mu\sqcap\mkern-7.5mu\sqcup\mkern1.5mu}T(\text{``state''})
   ]\mkern-3.3mu]
   =
   \left\{
      \begingroup\SmallColSep
      \begin{array}{r l}
         0
         &
         \mkern3mu
         ,
         \mkern12mu
         \sqcap\mkern-7.5mu\sqcup
         =
         \sqcap
         \\[3pt]
         1
         &
         \mkern3mu
         ,
         \mkern12mu
         \sqcap\mkern-7.5mu\sqcup
         =
         \sqcup
      \end{array}
      \endgroup   
   \right.
   \;\;\;\;  .
\end{equation}
\smallskip

\noindent As it can be observed, the functions $H{\mkern1.5mu\sqcap\mkern1.5mu}T(\text{``state''})$ and $H{\mkern1.5mu\sqcup\mkern1.5mu}T(\text{``state''})$ are generally not truth-functional: They return “false” and “true”, respectively, in any state of the coin, including its superposed state “heads up and tails up” wherein the statements $H$ and $T$ have no truth value at all.\\

\noindent Unlike $H{\mkern1.5mu\sqcap\mkern-7.5mu\sqcup\mkern1.5mu}T(\text{``state''})$, the statemental function \text{\guillemotleft}If ``state'' is true, then “at rest” \emph{and/or} “heads up” is true\text{\guillemotright}, denoted by $R{\mkern1.5mu\sqcap\mkern-7.5mu\sqcup\mkern1.5mu}H(\text{``state''})$, \emph{has no analytic proof supporting its either truth value}. This is because the instrumentalist description of quantum mechanics does not contain mathematical synonyms for compound categorical properties like “at rest” \emph{and} “heads up”, “at rest” \emph{or} “heads up”, “at rest” \emph{and} “tails up” and so forth. Put differently, in this description, it is impossible to assign Hilbert subspaces to compound categorical properties of this kind in such a way that replacing the assigned subspaces by the different ones would lead to contradiction.\\

\noindent To be sure, let us take up again $H{\mkern1.5mu\sqcap\mkern-7.5mu\sqcup\mkern1.5mu}T(\text{``state''})$. In addition, let us stipulate that “state” is given and so the statemental functions are equal in meaning to the corresponding statements.\\

\noindent By the implied significance of the subspace $\mathcal{E}_{H{\mkern1.5mu\sqcap\mkern1.5mu}T}$ representing the statement $H{\mkern1.5mu\sqcap\mkern1.5mu}T$, both the observable $\hat{O}_{H}$ and the observable $\hat{O}_{T}$ must leave every vector of $\mathcal{E}_{H{\mkern1.5mu\sqcap\mkern1.5mu}T}$ invariant. So, for all $|\Psi\rangle\in\mathcal{E}_{H{\mkern1.5mu\sqcap\mkern1.5mu}T}$, it must be true that $\hat{O}_{H}\hat{O}_{T}|\Psi\rangle=|\Psi\rangle$ as well as $\hat{O}_{T}\hat{O}_{H}|\Psi\rangle=|\Psi\rangle$. Providing $\hat{O}_{H{\mkern1.5mu\sqcap\mkern1.5mu}T}$ is the observable encoding $H{\mkern1.5mu\sqcap\mkern1.5mu}T$, the last means that $\hat{O}_{H{\mkern1.5mu\sqcap\mkern1.5mu}T}=\hat{O}_{H}\hat{O}_{T}=\hat{O}_{T}\hat{O}_{H}$. At this point, recall that the statements $H$ and $T$ are the negations of each other. This implies $\hat{O}_{H}=\hat{1}-\hat{O}_{T}$ and in turn $\hat{O}_{H{\mkern1.5mu\sqcap\mkern1.5mu}T}=0$. That is why $\mathcal{E}_{H{\mkern1.5mu\sqcap\mkern1.5mu}T}=\{0\}$.\\

\noindent Consider the subspace $\mathcal{E}_{H{\mkern1.5mu\sqcup\mkern1.5mu}T}$ representing the statement $H{\mkern1.5mu\sqcup\mkern1.5mu}T$. Since every vector $|\Psi_{H}\rangle\in\mathcal{E}_{H}$ is orthogonal to every vector $|\Psi_{T}\rangle\in\mathcal{E}_{T}$, namely, $\langle\Psi_{H}|\Psi_{T}\rangle=\langle\Psi_{H}|\hat{O}_{H}\hat{O}_{T}|\Psi_{T}\rangle=0$, an arbitrary vector $|\Psi\rangle\in\mathcal{E}_{H{\mkern1.5mu\sqcup\mkern1.5mu}T}$ can be uniquely decomposed into the sum of two pieces as $|\Psi\rangle=c_{H}|\Psi_{H}\rangle+c_{T}|\Psi_{T}\rangle$, where $c_{H}|\Psi_{H}\rangle=\hat{O}_{H}|\Psi\rangle$ and $c_{T}|\Psi_{T}\rangle=\hat{O}_{T}|\Psi\rangle$. Thus, $(\hat{O}_{H}+\hat{O}_{T})|\Psi\rangle=|\Psi\rangle\in\mathcal{E}_{H{\mkern1.5mu\sqcup\mkern1.5mu}T}$, which means that the observable encoding the statement $H{\mkern1.5mu\sqcup\mkern1.5mu}T$ is $\hat{O}_{H{\mkern1.5mu\sqcup\mkern1.5mu}T}=\hat{O}_{H}+\hat{O}_{T}=1$. For this reason, $\mathcal{E}_{H{\mkern1.5mu\sqcup\mkern1.5mu}T}=\mathcal{H}$.\\

\noindent As the statements $R$ and $H$ are not the negations of each other, $\hat{O}_{R}\hat{O}_{H}{\neq}0$. Consequently, eigenvectors $|\Psi_{R}\rangle\in\mathcal{E}_{R}$ and $|\Psi_{H}\rangle\in\mathcal{E}_{H}$ of the observables $\hat{O}_{R}$ and $\hat{O}_{H}$ are not orthogonal: $\langle\Psi_{R}|\Psi_{H}\rangle=\langle\Psi_{R}|\hat{O}_{R}\hat{O}_{H}|\Psi_{H}\rangle{\mkern2mu\neq\mkern4mu}0$. Hence, an arbitrary vector $|\Psi\rangle$ of the subspace $\mathcal{E}_{R{\mkern1.5mu\sqcup\mkern1.5mu}H}$ representing the statement $R{\mkern1.5mu\sqcup\mkern1.5mu}H$ cannot be uniquely decomposed into the sum $\hat{O}_{R}|\Psi\rangle+\hat{O}_{H}|\Psi\rangle$. Accordingly, for some (nonzero) $|\Psi\rangle\in\mathcal{E}_{R{\mkern1.5mu\sqcup\mkern1.5mu}H}$, it holds true that $(\hat{O}_{R}+\hat{O}_{H})|\Psi\rangle\notin\mathcal{E}_{R{\mkern1.5mu\sqcup\mkern1.5mu}H}$. Therefore, the observable $\hat{O}_{R{\mkern1.5mu\sqcup\mkern1.5mu}H}$ encoding $R{\mkern1.5mu\sqcup\mkern1.5mu}H$ cannot be the sum of $\hat{O}_{R}$ and $\hat{O}_{H}$.\\

\enlargethispage{\baselineskip}
\enlargethispage{\baselineskip}
\noindent Likewise, the observable $\hat{O}_{R{\mkern1.5mu\sqcap\mkern1.5mu}H}$ encoding the statement $R{\mkern1.5mu\sqcap\mkern1.5mu}H$ cannot be the product of $\hat{O}_{R}$ and $\hat{O}_{H}$. That is, for some (nonzero) $|\Psi\rangle\in\mathcal{E}_{R{\mkern1.5mu\sqcap\mkern1.5mu}H}$, it holds true that $(\hat{O}_{R}\hat{O}_{H})|\Psi\rangle\notin\mathcal{E}_{R{\mkern1.5mu\sqcap\mkern1.5mu}H}$ as well as $(\hat{O}_{H}\hat{O}_{R})|\Psi\rangle\notin\mathcal{E}_{R{\mkern1.5mu\sqcap\mkern1.5mu}H}$.\\

\noindent Clearly, the inequalities demanded by the quantum instrumentalist description as conditions for the observables $\hat{O}_{R{\mkern1.5mu\sqcup\mkern1.5mu}H}$ and $\hat{O}_{R{\mkern1.5mu\sqcap\mkern1.5mu}H}$, namely,\smallskip

\begin{equation} \label{COND1} %{Eq.59}
   \hat{O}_{R{\mkern1.5mu\sqcup\mkern1.5mu}H}
   \neq
   \hat{O}_{R}
   +
   \hat{O}_{H}
   \;\;\;\;  ,
\end{equation}
\\[-36pt]

\begin{equation} \label{COND2} %{Eq.60}
   \begingroup\SmallColSep
   \begin{array}{l}
      \hat{O}_{R{\mkern1.5mu\sqcap\mkern1.5mu}H}
      \neq
      \hat{O}_{R}
      \hat{O}_{H}
      \\[6pt]
      \hat{O}_{R{\mkern1.5mu\sqcap\mkern1.5mu}H}
      \neq
      \hat{O}_{H}
      \hat{O}_{R}
   \end{array}
   \endgroup   
   \;\;\;\;  ,
\end{equation}
\smallskip

\noindent are not able to define the corresponding Hilbert subspaces in a manner that $\mathcal{E}_{R{\mkern1.5mu\sqcup\mkern1.5mu}H}$ and $\mathcal{E}_{R{\mkern1.5mu\sqcap\mkern1.5mu}H}$ are irreplaceable.\\

\noindent For example, since the observable $\hat{O}_{\top}$ is not equal to the sum $\hat{O}_{R}+\hat{O}_{H}$ and neither the product $\hat{O}_{R}\hat{O}_{H}$ nor the product $\hat{O}_{H}\hat{O}_{R}$ is the observable $\hat{O}_{\bot}$, the identical subspace $\mathcal{H}$ and the intersection $\mathcal{E}_{R}\cap\mathcal{E}_{H}=\{0\}$ can be assigned to $\mathcal{E}_{R{\mkern1.5mu\sqcup\mkern1.5mu}H}$ and $\mathcal{E}_{R{\mkern1.5mu\sqcap\mkern1.5mu}H}$, respectively. Then again, the said assignment can be denied without contradiction: Subspaces $\mathcal{E}^{\prime}\subset\mathcal{H}$ and $\mathcal{E}^{\prime\prime}\neq\{0\}$ may also be assigned, respectively, to $\mathcal{E}_{R{\mkern1.5mu\sqcup\mkern1.5mu}H}$ and $\mathcal{E}_{R{\mkern1.5mu\sqcap\mkern1.5mu}H}$, on condition that their corresponding observables $\hat{O}^{\prime}$ and $\hat{O}^{\prime\prime}$ meets the inequalities (\ref{COND1}) and (\ref{COND2}).\\

\noindent As appears, neither of the subspaces of $\mathcal{H}$ is the mathematical synonym for the compound categorical property “at rest” \emph{and} “heads up”. The same applies to the compound categorical property “at rest” \emph{or} “heads up”. Hence, the analytic proof of the truth or falsity is not available for the statements $R{\mkern1.5mu\sqcup\mkern1.5mu}H$ and $R{\mkern1.5mu\sqcap\mkern1.5mu}H$, which makes them truth-valueless within the alternative model of reasoning.\\

\noindent In a succinct form, the valuation for arbitrary statemental functions $S_1{\mkern1.5mu\sqcap\mkern-7.5mu\sqcup\mkern1.5mu}S_2(\text{``state''})$ can be written as:\smallskip

\begin{equation}  %{Eq.61}
   [\mkern-3.3mu[
      S_1{\mkern1.5mu\sqcap\mkern-7.5mu\sqcup\mkern1.5mu}S_2(\text{``state''})
   ]\mkern-3.3mu]
   =
   \left\{
      \begingroup\SmallColSep
      \begin{array}{r l}
         1
         &
         \mkern3mu
         ,
         \mkern12mu
         [\mkern-3.3mu[
            \sigma_{S_1{\mkern1.5mu\sqcap\mkern-7.5mu\sqcup\mkern1.5mu}S_2}(|\Psi\rangle)
         ]\mkern-3.3mu]
         =
         1
         \\[4pt]
         0
         &
         \mkern3mu
         ,
         \mkern12mu
         [\mkern-3.3mu[
           \alpha_{S_1{\mkern1.5mu\sqcap\mkern-7.5mu\sqcup\mkern1.5mu}S_2}(|\Psi\rangle)
         ]\mkern-3.3mu]
         =
         1
         \\[4pt]
         \textnormal{NULL}
         &
         \mkern3mu
         ,
         \mkern12mu
         \left\{
            \begingroup\SmallColSep
            \begin{array}{l}
               [\mkern-3.3mu[
                  \sigma_{S_1{\mkern1.5mu\sqcap\mkern-7.5mu\sqcup\mkern1.5mu}S_2}(|\Psi\rangle)
               ]\mkern-3.3mu]
               =
               [\mkern-3.3mu[
                  \alpha_{S_1{\mkern1.5mu\sqcap\mkern-7.5mu\sqcup\mkern1.5mu}S_2}(|\Psi\rangle)
               ]\mkern-3.3mu]
               =
               0
               \\[3pt]
               \sigma_{S_1{\mkern1.5mu\sqcap\mkern-7.5mu\sqcup\mkern1.5mu}S_2}(|\Psi\rangle)
               \mkern4mu
               \textnormal{and}
               \mkern4mu
               \alpha_{S_1{\mkern1.5mu\sqcap\mkern-7.5mu\sqcup\mkern1.5mu}S_2}(|\Psi\rangle)
               \mkern4mu
               \textnormal{do not exist}
            \end{array}
            \endgroup
      \right.
      \end{array}
      \endgroup   
   \right.
   \;\;\;\;  ,
\end{equation}
\smallskip

\noindent where $\sigma_{S_1{\mkern1.5mu\sqcap\mkern-7.5mu\sqcup\mkern1.5mu}S_2}(|\Psi\rangle)$ and $\alpha_{S_1{\mkern1.5mu\sqcap\mkern-7.5mu\sqcup\mkern1.5mu}S_2}(|\Psi\rangle)$ stand for the functions’ mathematical synonyms and antonyms, respectively.\\

\noindent To equip the alternative model of reasoning with an implication operation, the valuation analogous to the above can be proposed for the statemental function $S_1{\to}S_2(\text{``state''})$, where the mathematical synonym and antonym for this function, i.e., $\sigma_{S_1{\to}S_2}(|\Psi\rangle)$ and $\alpha_{S_1{\to}S_2}(|\Psi\rangle)$ in that order, are defined as follows:\smallskip

\begin{equation}  %{Eq.62}
   \sigma_{S_1{\to}S_2}(|\Psi\rangle)
   =
   \Big\{
      |\Psi\rangle
      \in
      \mathcal{H}
      \Big|
         |\Psi\rangle
         \in
         \mathcal{E}_{{\mkern1.5mu\neg}S_1{\mkern1.5mu\sqcup\mkern1.5mu}S_2}
      \Big.
   \Big\}
   \;\;\;\;  ,
\end{equation}
\\[-36pt]

\begin{equation}  %{Eq.63}
   \alpha_{S_1{\to}S_2}(|\Psi\rangle)
   =
   \Big\{
      |\Psi\rangle
      \in
      \mathcal{H}
      \Big|
         |\Psi\rangle
         \in
         \mathcal{E}_{S_1{\mkern1.5mu\sqcap\mkern1.5mu}{\neg}S_2}
      \Big.
   \Big\}
   \;\;\;\;  .
\end{equation}
\smallskip

\noindent According to the said definition,\smallskip

\begin{equation}  %{Eq.64}
   [\mkern-3.3mu[
     H{\mkern1.5mu\to\mkern3mu}T(\text{``state''})
   ]\mkern-3.3mu]
   =
   \left\{
      \begingroup\SmallColSep
      \begin{array}{r l}
         1
         &
         \mkern3mu
         ,
         \mkern12mu
         [\mkern-3.3mu[
            \mkern1.5mu|\Psi\rangle\in\mathcal{E}_{T}
         ]\mkern-3.3mu]
         =
         1
         \\[3pt]
         0
         &
         \mkern3mu
         ,
         \mkern12mu
         [\mkern-3.3mu[
            \mkern1.5mu|\Psi\rangle\in\mathcal{E}_{H}
         ]\mkern-3.3mu]
         =
         1
         \\[3pt]
         \textnormal{NULL}
         &
         \mkern3mu
         ,
         \mkern12mu
         [\mkern-3.3mu[
            \mkern1.5mu|\Psi\rangle\in\mathcal{E}_{T}
         ]\mkern-3.3mu]
         =
         [\mkern-3.3mu[
            \mkern1.5mu|\Psi\rangle\in\mathcal{E}_{H}
         ]\mkern-3.3mu]
         =
         0
      \end{array}
      \endgroup   
   \right.
   \;\;\;\;  .
\end{equation}
\smallskip

\noindent Thus, the statement $H{\mkern1.5mu\to\mkern3mu}T(\text{``heads up''})$ is false, the statement $H{\mkern1.5mu\to\mkern3mu}T(\text{``tails up''})$ is true, and the statement $H{\mkern1.5mu\to\mkern3mu}T(\text{``heads up and tails up''})$ has no truth value at all.\\

\noindent For comparison,

\begin{equation}  %{Eq.65}
   [\mkern-3.3mu[
     R{\mkern1.5mu\to\mkern3mu}H(\text{``state''})
   ]\mkern-3.3mu]
   =
  \textnormal{NULL}
   \;\;\;\;  ,
\end{equation}
\\[-36pt]

\begin{equation}  %{Eq.66}
   [\mkern-3.3mu[
     R{\mkern1.5mu\to\mkern3mu}T(\text{``state''})
   ]\mkern-3.3mu]
   =
  \textnormal{NULL}
   \;\;\;\;  ,
\end{equation}
\smallskip

\noindent because none of $\mathcal{E}_{{\mkern1.5mu\neg}R{\mkern1.5mu\sqcup\mkern1.5mu}H}$, $\mathcal{E}_{{\mkern1.5mu\neg}R{\mkern1.5mu\sqcup\mkern1.5mu}T}$, $\mathcal{E}_{R{\mkern1.5mu\sqcap\mkern1.5mu}{\mkern1.5mu\neg}H}$, and $\mathcal{E}_{R{\mkern1.5mu\sqcap\mkern1.5mu}{\mkern1.5mu\neg}T}$ can be assigned a Hilbert subspace in an irreplaceable manner. In line with that, the statemental function \text{\guillemotleft}If the coin’s state is “state”, then the categorical property “at rest” of the coin implies its categorical property ``heads up'' (``tails up'')\text{\guillemotright} must be declared to be valueless without consideration for the state.\\

\section{The alternative model of reasoning vs. quantum logic}  %{<-------------------------------------------------------------------------------------------------Section IX}

\noindent Let us examine the differences between the alternative model of reasoning outlined in the previous section (and henceforth called AMR for brevity) and quantum logic (which will henceforth be called QL), the most elaborated version of non-classical logic to date designed to analyze and evaluate quantum statements (i.e., the ones that concern quantum phenomena).\\

\noindent First and foremost, QL is an interpretation of quantum mechanics. Being as such, QL adds an extra postulate to the instrumentalist description of quantum theory. Explicitly, QL postulates truth values for quantum statements like \text{\guillemotleft}If ``state'' is true, then ``$s_1$”$\ast$``$s_2$” is true\text{\guillemotright}, where ``$s_1$”$\ast$``$s_2$” is a compound that merges incompatible categorical properties (such as ``at rest” and ``heads up”). All the rest claimed by QL – e.g., tying quantum statements to closed linear subspaces of a Hilbert space and representing logical operations on quantum statements as lattice-theoretic operations on those subspaces – serves to supplement the said postulate \cite{Dickson}.\\

\noindent However, not only is this postulate outside of the quantum instrumental description (which has no mathematical synonyms and antonyms for compounds that bring together incompatible categorical properties), but also it is unfalsifiable, i.e., it is unable – even hypothetically – to come in conflict with observation: According to quantum theory, incompatible properties are unobservable all together. Thus, the main postulate of quantum logic is unphysical.\\

\noindent By contrast, AMR does not lead to a new interpretation of quantum mechanics for it does not decide a truth value of a statement outside the quantum instrumental description.\\

\noindent Furthermore, QL asserts simultaneous definability of all quantum statements \cite{Dickson}. To make this assertion consistent with the Kochen-Specker theorem \cite{Kochen}, which rules out the idea that every quantum statement has a definite truth value, one is obliged to introduce a new property of a statement such as \emph{meaningfulness} (or \emph{noncontextuality}). This property is not the same as a truth value; to be more exact, a quantum statement can be meaningful only within \emph{a certain context}, i.e., a maximally possible set of quantum statements encoded by mutually compatible observables on the Hilbert space of a physical system (such as the set of the statements $\{H,T\}$) \cite{Griffiths}. As a result, one may argue that the Kochen-Specker theorem excludes only the ability to simultaneously assign meaningful (noncontextual) definite truth values to all quantum statements \cite{Abbott}.\\

\noindent That inconsistency is absent in AMR: As a matter of fact, the Kohen-Speckler theorem proves the impossibility of an arbitrary interpretation of all the “atoms” regarding a quantum system (which results in that some “atoms” have no definite truth value). Therefore, this theorem can be regarded as the collateral consequence of the first principle of AMR that demands proof for assigning a truth value to each “atom” (since in a certain context such proof is not present for some “atoms”, they are valueless in that context).\\

\noindent But most importantly, QL struggles to explain how uncertainties (and, thus, probabilities as their measures) can emerge in the context of (non-classical) propositional logic, i.e., a deterministic model of reasoning which involves only certain (i.e., confident and assured) truths and inferences.\\

\noindent To elucidate this point, consider the argument:\\[-22pt]

\[
\begin{array}{l l}
   \text{\texttt{Premise 1}}
   &
   \text{If a state of the coin is not a mixture of other states, then every categorical}
   \\[-1pt]
   \qquad
   &
   \text{property of the coin is determined.}
   \\[6pt]
   \text{\texttt{Premise 2}}
   &
   \text{The state of the coin is not a mixture of other states.}
   \\[6pt]
   \text{\texttt{Conclusion}}
   &
   \text{Every categorical property of the coin is determined.}
\end{array}
\]
\\[-17pt]

\noindent This argument can be formalized by the following expressions in predicate logic:\smallskip

\begin{equation} \label{ } %{Eq.67}
   \begingroup\SmallColSep
   \begin{array}{c c l}
      \texttt{P1}
      &
      \qquad
      &
      \text{``state''}\notin\mathbb{M}\Longrightarrow\forall{S(\text{``state''})}\in\mathbb{S}(\text{``state''})\Big([\mkern-3.3mu[S(\text{``state''}) ]\mkern-3.3mu]\in\{0,1\}\Big)
      \\[4pt]
      \texttt{P2}
      &
      \qquad
      &
      \text{``state''}\notin\mathbb{M}
      \\[4pt]
      \mathtt{\therefore}
      &
      \qquad
      &
      \forall{S(\text{``state''})}\in\mathbb{S}(\text{``state''})\Big([\mkern-3.3mu[S(\text{``state''}) ]\mkern-3.3mu]\in\{0,1\}\Big)
   \end{array}
   \endgroup
   \;\;\;\;  ,
\end{equation}
\smallskip

\noindent where $\texttt{P1}$, $\texttt{P2}$ and $\mathtt{\therefore}$ stand for the premises and the conclusion, respectively, $\mathbb{M}$ denotes the set of all states of the coin that are mixtures of other states, $S(\text{``state''})$ is the shorthand for \text{\guillemotleft}If the state of a physical system is “state”, then the system’s categorical property is “$s$”\text{\guillemotright}, and $\mathbb{S}(\text{``state''})$ represents the set of all quantum statements (relating to “state”). The argument in question is valid because it utilizes the rule of \emph{modus ponens} (i.e., the premises entail the conclusion).\\

\noindent QL declares the simultaneous well-definedness (unambiguousness) of all quantum statements, which implies that any quantum statement is always firmly resolved – i.e., either true or false – in a state that is not a mixture of other states. Expressly, the following sentence is true in the frame of QL:\smallskip

\begin{equation}  %{Eq.68}
   \forall\mkern2mu\text{``state''}\notin\mathbb{M}
   \Big[
      \forall{S(\text{``state''})}\in\mathbb{S}(\text{``state''})\Big([\mkern-3.3mu[S(\text{``state''}) ]\mkern-3.3mu]\in\{0,1\}\Big)
   \Big]
   \;\;\;\;  .
\end{equation}
\smallskip

\noindent Hence, the premise 1 is true in QL. Also, let us suppose that the premise 2 is true. Then, since the argument is valid and both premises are true, its conclusion must be true as well.\\

\noindent On the other hand, the true conclusion of the argument contradicts quantum theory. By way of illustration, in the superposed state ”heads up and tails up” (which is not a mixture of other states), the probability of the categorical property of the coin being “heads up” is different from both 1 and 0. Providing the dispersion-free probabilities (i.e., the probability values 1 and 0) correspond to the truth values “true” and “false”, respectively, the last means that\smallskip

\begin{equation}  %{Eq.69}
   [\mkern-3.3mu[H(\text{``heads up and tails up''})]\mkern-3.3mu]
   \notin\{0,1\}
   \;\;\;\;  .
\end{equation}
\smallskip

\noindent Thus, \emph{the premises of QL cannot logically lead to the emergence of dispersion-included probabilities} (i.e., probability values that are greater than 0 and less than 1).\\

\noindent By contrast, the premise 1 is false in AMR. This is due to the fact that to assign a truth value to the quantum statement under study, AMR requires analytic proof, which may possibly not exist for the statement. As a consequence, it is not the case that, given any statemental function $S(\text{``state''})$, that function has a truth value. Symbolically, this can be presented as\smallskip

\begin{equation}  %{Eq.70}
   [\mkern-3.3mu[S(\text{``state''}) ]\mkern-3.3mu]
   \in
   \{
      0
      ,
      1
      ,
      \textnormal{NULL}
   \}
   \mkern4mu
   \Longrightarrow
   \mkern4mu
   \neg
   \forall{S(\text{``state''})}\in\mathbb{S}(\text{``state''})\Big([\mkern-3.3mu[S(\text{``state''}) ]\mkern-3.3mu]\in\{0,1\}\Big)
   \;\;\;\;  .
\end{equation}
\smallskip

\noindent Hence, even though the argument in question remains valid, the falsity of the premise 1 makes the conclusion of the argument false. Therefore, the sentence\smallskip

\begin{equation}  %{Eq.71}
      \exists{S(\text{``state''})}\in\mathbb{S}(\text{``state''})
      \Big(
         \Pr\mkern-1.5mu\left[S(\text{``state''})\right]
         \notin\{0,1\}
      \Big)
   \;\;\;\;  ,
\end{equation}
\smallskip

\noindent which affirms that the set of all quantum statements $\mathbb{S}(\text{``state''})$ does not admit only dispersion-free probabilities, must be true. This truth matches the consequence of Gleason's theorem \cite{Gleason}, which is used to put dispersion-included probabilities in QL (in order to get them out) \cite{Wilce}.\\

\noindent On that account, one may state that unlike QL, \emph{AMR already contain some probabilistic concept}. That is, within the frame of AMR probabilities are not introduced but conserved: They are passed from the premise(s) to the conclusion.\\

\noindent As a means to study dispersion-included probabilities (instead of dispersion-free probabilities, i.e., truth values), one can replace truth assignment with probabilistic semantics.\\

\noindent To accomplish this, let us first briefly dwell on the way, in which the objective truth values “true” and ``false'' behave towards the epistemic predicates “\emph{verified}” and “\emph{falsified}” (which is synonymous with “\emph{refuted}”). On the one hand, “true” and “false” cannot be identified with “verified” and “falsified”, for a true (false) statement may be verified (falsified) at one time and not at another. But on the other hand, if a statement is ever accepted as verified (falsified), then from that time onward this statement can be said to be a true (false) one \cite{Varzi}.\\

\noindent Suppose that at some time, a statement is verified (falsified) by an experience (e.g., observation or experiment). Understandably, were the statement to have a definite truth value prior to the experience, the affirmative outcome of the verification (equally, the negative outcome of the refutation) would be either \emph{certain} or \emph{impossible}.\\

\noindent Take as an example the true statement \text{\guillemotleft}2+2=4\text{\guillemotright}. Without doubt, the affirmative outcome of the verification of \text{\guillemotleft}2+2=4\text{\guillemotright} is certain (as much as the negative outcome of the refutation of \text{\guillemotleft}2+2=4\text{\guillemotright}). By contrast, the positive result ensuing from the verification of the false statement \text{\guillemotleft}2+2=5\text{\guillemotright} is impossible (the same as the negative result following the action of proving \text{\guillemotleft}2+2=5\text{\guillemotright} to be false).\\

\noindent Now take the statement \text{\guillemotleft}P=NP\text{\guillemotright} asserting that every NP problem, i.e., the one that can be verified quickly, is also a P problem, i.e., the one that can be solved quickly \cite{Cormen}. As of today, neither proof of truth nor proof of falsity is present for this statement and so it has no truth value up to now, according to AMR. Nevertheless, it is possible that someday this statement will be proved true as much as it is possible that it will be proved false. In other words, the affirmative (negative) outcome of the verification (refutation) of the statement \text{\guillemotleft}P=NP\text{\guillemotright} is neither certain nor impossible.\\

\noindent The above examples can be presented symbolically as follows\smallskip

\begin{equation}  %{Eq.72}
   \Pr
   \left[
      \text{\guillemotleft}2+2=4\text{\guillemotright}
   \right]
   =
   1
   \;\;\;\;  ,
\end{equation}
\\[-36pt]

\begin{equation}  %{Eq.73}
   \Pr
   \left[
      \text{\guillemotleft}2+2=5\text{\guillemotright}
   \right]
   =
   0
   \;\;\;\;  ,
\end{equation}
\\[-36pt]

\begin{equation}  %{Eq.74}
   \Pr
   \left[
      \text{\guillemotleft}\mathrm{P}=\mathrm{NP}\text{\guillemotright}
   \right]
   \notin
   \{0,1\}
   \;\;\;\;  ,
\end{equation}
\smallskip

\noindent where $\Pr[\cdot]$ denotes the probability of the affirmative outcome of the verification (or the negative outcome of the refutation) of a statement (symbolized by the interpunct ``$\cdot$''), while the probability values 1 and 0 signify certainty and impossibility in that order.\\

\noindent This gives a reason to introduce probabilistic semantics for any statemental function $S(\text{``state''})$ in the following way:\smallskip

\begin{equation}  %{Eq.75}
   \begingroup
   \begin{array}{r c c l}
      \Pr:
      &
      \mathbb{S}(\text{``state''})
      &
      \to
      &
      [0,1]
      \\[5pt]
      \hfill
      &
      S(\text{``state''})
      &
      \mapsto
      &
      \Pr\left[S(\text{``state''})\right]
   \end{array}
   \endgroup
   \;\;\;\;  ,
\end{equation}
\smallskip

\noindent where $\mathbb{S}(\text{``state''})$ is the set of statemental functions, $\Pr\left[S(\text{``state''})\right]$ is the image of $S(\text{``state''})$ under $\Pr$ defined by the expression\smallskip

\begin{equation}  %{Eq.76}
   \Pr\left[S(\text{``state''})\right]
   =
   \left\{
      \begingroup\SmallColSep
      \begin{array}{r l}
         1
         &
         \mkern3mu
         ,
         \mkern12mu
         [\mkern-3.3mu[
            \sigma_{S}(|\Psi\rangle)
         ]\mkern-3.3mu]
         =
         1
         \\[4pt]
         0
         &
         \mkern3mu
         ,
         \mkern12mu
         [\mkern-3.3mu[
            \alpha_{S}(|\Psi\rangle)
         ]\mkern-3.3mu]
         =
         1
         \\[4pt]
         r\in(0,1)
         &
         \mkern3mu
         ,
         \mkern12mu
         [\mkern-3.3mu[
            \sigma_{S}(|\Psi\rangle)
         ]\mkern-3.3mu]
         =
         [\mkern-3.3mu[
            \alpha_{S}(|\Psi\rangle)
         ]\mkern-3.3mu]
         =
         0   
         \\[4pt]
         \textnormal{NULL}
         &
         \mkern3mu
         ,
         \mkern12mu
         \textnormal{neither}
         \mkern5mu
         \sigma_{S}(|\Psi\rangle)
         \mkern5mu
         \textnormal{nor}
         \mkern5mu
         \alpha_{S}(|\Psi\rangle)
         \mkern5mu
         \textnormal{exists}
      \end{array}
      \endgroup
   \right.
   \;\;\;\;  ,
\end{equation}
\smallskip

\noindent in which $\sigma_{S}(|\Psi\rangle)$ and $\alpha_{S}(|\Psi\rangle)$ are the mathematical synonym and antonym, correspondingly, for $S(\text{``state''})$.\\

\noindent For example, in the state “heads up and tails up” represented by the vector $|\mkern-0.5mu\Psi_{\mkern-0.5mu{R}}\mkern-0.5mu\rangle=c_{H} |\mkern-0.5mu\Psi_{\mkern-0.5mu{H}}\mkern-0.5mu\rangle+c_{T}|\mkern-0.5mu\Psi_{\mkern-0.5mu{T}}\mkern-0.5mu\rangle$, both the synonym and the antonym for the function $H(\text{``state''})$ are false; for this reason, the probabilistic valuation of the statement $H(\text{``heads up and tails up''})$ admits neither value 1 nor value 0:\smallskip

\begin{equation}  %{Eq.77}
   \Pr\left[H(\text{``heads up and tails up''})\right]
   \in
   (0,1)
   \;\;\;\;  .
\end{equation}
\smallskip

\noindent The same holds for the function $T(\text{``state''})$.\\

\noindent Because the mathematical synonyms for the statemental functions $H{\mkern1.5mu\sqcup\mkern1.5mu}T(\text{``state''})$ and $H{\mkern1.5mu\sqcap\mkern1.5mu}T(\text{``state''})$ are the universal set (consisting of all admissible states of the coin, including the superposed state ``heads up and tails up'') and its complement, the empty set, respectively, the probabilistic valuation for these functions must be 1 and 0 in that order:\smallskip

\begin{equation}  %{Eq.78}
   [\mkern-3.3mu[\sigma_{H{\mkern1.5mu\sqcup\mkern1.5mu}T}(|\Psi\rangle)]\mkern-3.3mu]
   =
   1
   :
   \mkern12mu
   \Pr\left[H{\mkern1.5mu\sqcup\mkern1.5mu}T(\text{``state''})\right]
   =
   1
   \;\;\;\;  ,
\end{equation}
\\[-36pt]

\begin{equation}  %{Eq.79}
   [\mkern-3.3mu[\alpha_{H{\mkern1.5mu\sqcap\mkern1.5mu}T}(|\Psi\rangle)]\mkern-3.3mu]
   =
   1
   :
   \mkern12mu
   \Pr\left[H{\mkern1.5mu\sqcap\mkern1.5mu}T(\text{``state''})\right]
   =
   0
   \;\;\;\;  .
\end{equation}
\smallskip

\noindent Hence, in accordance with the addition law of probability, the following holds true:\smallskip

\begin{equation}  %{Eq.80}
   \Pr\left[H{\mkern1.5mu\sqcup\mkern1.5mu}T(\text{``state''})\right]
   =
   \Pr\left[H(\text{``state''})\right]
   +
   \Pr\left[T(\text{``state''})\right]
   =
   1
   \;\;\;\;  .
\end{equation}
\smallskip

\noindent As the coin is ideal, the functions $H$ and $T$ must have the equal degree of (un)certainty in the state “heads up and tails up”; thus\smallskip

\begin{equation}  %{Eq.81}
   \Pr\left[H(\text{``heads up and tails up''})\right]
   =
   \frac{1}{2}
   \;\;\;\;  .
\end{equation}
\smallskip

\noindent The above value coincides with the Born probability\smallskip

\begin{equation}  %{Eq.82}
   \left|
      \langle\Psi_{R}|\Psi_{H}\rangle
   \right|^2
   =
   \frac{1}{2}
   \;\;\;\;  ,
\end{equation}
\smallskip

\noindent where $|\mkern-0.5mu\Psi_{\mkern-0.5mu{R}}\mkern-0.5mu\rangle=\rfrac{1}{\sqrt{2}}|\mkern-0.5mu\Psi_{\mkern-0.5mu{H}}\mkern-0.5mu\rangle+\rfrac{1}{\sqrt{2}}|\mkern-0.5mu\Psi_{\mkern-0.5mu{T}}\mkern-0.5mu\rangle$ describes the superposed state of the coin in which its categorical property is an amount $\rfrac{1}{\sqrt{2}}$ “heads up” and an amount $\rfrac{1}{\sqrt{2}}$ “tails up”.\\

\noindent It may be worth noting – albeit, just in passing – that in the context of AMR, the probabilistic valuation of a statemental function like $R{\mkern1.5mu\sqcap\mkern1.5mu}H(\text{``state''})$ returns no value as much as the function’s truth valuation does. This can be explained by pointing out that the verification (refutation) of the statement $R{\mkern1.5mu\sqcap\mkern1.5mu}H$ in one or another state cannot be achieved due to uncommutativeness of the observables $\hat{O}_{R}$ and $\hat{O}_{H}$ (in another words, by virtue of the absence of the mathematical synonym and antonym for the function $R{\mkern1.5mu\sqcap\mkern1.5mu}H(\text{``state''})$ in the quantum instrumental description). Since the said verification (refutation) cannot happen, $\Pr\left[R{\mkern1.5mu\sqcap\mkern1.5mu}H(\text{``state''})\right]$, the probability of its affirmative (negative) outcome, is of no value. In consequence, the problem of joint probability for non-commuting observables \cite{Busch} does not exist in AMR.\\

\section{Concluding remarks}  %{<-------------------------------------------------------------------------------------------------Section X}

\noindent Let us make a sketch of what has been discussed in this paper.\\

\noindent The atomic statement \text{\guillemotleft}If “heads up and tails up” is true, then “heads up” is true\text{\guillemotright}, which asserts that the categorical property of the coin is “heads up” in the superposed state “heads up and tails up”, is not analytic. The reason is that the mathematical formalism of quantum theory does not define the meaning of the coordinating conjunction “and” in the noun phrase “heads up and tails up”. As a result, the instrumental description of quantum theory provides no proof for either truth value of the abovementioned statement.\\

\noindent But then again, if a logical system – for instance, propositional logic or some modification of it – is justifiable in the physical universe, each “atom” concerning quantum phenomena may be interpreted freely, in particular, without being controlled or restricted by the laws of quantum theory. So, while the quantum instrumental description has no means to interpret the statement \text{\guillemotleft}If “heads up and tails up” is true, then “heads up” is true\text{\guillemotright}, it may be assigned a truth value in accordance with logic. The only conclusion, which is possible to draw from this, is that a truth assignment to the said statement should be sought outside the quantum instrumental description, i.e., it can be added to this description as an extra postulate.\\

\noindent This is how the problem of interpretations of quantum mechanics arises.\\

\noindent Suppose that the extra postulate is falsifiable. Because it does not belong to the set of basic postulates of quantum mechanics, an interpretation brought into existence by the extra postulate of this kind has the potential to produce predictions different from the ones of the mathematical formalism of quantum theory. However, at least at present there is no positive experimental evidence that the quantum formalism is not quantitatively valid.\\

\noindent On the other hand, if the extra postulate is unfalsifiable, i.e., the one that cannot be negated by future experiments, then an interpretation built on it will have no merit of being testable. Such an interpretation should be regarded as uncriticizable.\\

\noindent A way around the impasse over the problem of interpretations of quantum mechanics is to reject the implicit assumption, on which every logical system is based, to wit: a free (not based on proof) interpretation of the “atoms”.\\

\noindent Thus, in the alternative model of reasoning, mathematical truths, synonymous and antonymous with the “atom”, are proofs witnessing its truth and falsity, respectively. On the other hand, those mathematical synonyms and antonyms are identified with the quantum instrumental description. Hence, if the said description shows that the synonyms and antonyms are all false, the quantum “atom” bears no truth value. For that reason, the statement \text{\guillemotleft}If “heads up and tails up” is true, then “heads up” is true\text{\guillemotright} has no truth value. So, neither has the statement \text{\guillemotleft}If “heads up and tails up” is true, then “tails up” is true\text{\guillemotright}. In this way, a valuation of such statements need not be added as an extra postulate to the mathematical formalism of quantum theory.\\

\noindent Together with that, the quantum instrumental description does provide proof for the truth value of the compound statemental function \text{\guillemotleft}If ``state'' is true, then “heads up” \emph{and} “tails up” is true\text{\guillemotright} as well as the function \text{\guillemotleft}If ``state'' is true, then “heads up” \emph{or} “tails up” is true\text{\guillemotright}, denoted by $H{\mkern1.5mu\sqcap\mkern1.5mu}T(\text{``state''})$ and $H{\mkern1.5mu\sqcup\mkern1.5mu}T(\text{``state''})$ correspondingly. Specifically, those functions are false and true, in that order, in any admissible “state”, including ``heads up and tails up''.\\

\noindent In that respect, “quantum fundamentalism” and “macroscopic realism” are not contradictory within the framework of the alternative model of reasoning. This can be presented as\smallskip

\begin{equation}  %{Eq.83}
   [\mkern-3.3mu[
      H{\mkern1.5mu\sqcap\mkern1.5mu}T(\text{``before''})
   ]\mkern-3.3mu]
   =
   [\mkern-3.3mu[
      H{\mkern1.5mu\sqcap\mkern1.5mu}T(\text{``after''})
   ]\mkern-3.3mu]
   =
   0
   \;\;\;\;  ,
\end{equation}
\\[-36pt]

\begin{equation}  %{Eq.84}
   [\mkern-3.3mu[
      H{\mkern1.5mu\sqcup\mkern1.5mu}T(\text{``before''})
   ]\mkern-3.3mu]
   =
   [\mkern-3.3mu[
      H{\mkern1.5mu\sqcup\mkern1.5mu}T(\text{``after''})
   ]\mkern-3.3mu]
   =
   1
   \;\;\;\;  ,
\end{equation}
\smallskip

\noindent where “before” and “after” act for the states before and after observation (for example, “before” = “heads up and tails up” and “after” = “heads up”).\\

\noindent It is fairly straightforward to generalize what has been discussed above.\\

\noindent Suppose that $S_{i}(\text{``state''})$, where $i\in\{1,\dots,N\}$, is a statemental function pertaining to the act of observing an arbitrary physical system in the state denoted by “state” (in the case of the ideal coin, $N=2$ and $\{S_1,S_2\}=\{H,T\}$). The truth of the statement $S_{i}(\text{``before''})$ represents a potential outcome of observation, while the truth of the statement $S_{i}(\text{``after''})$ stands in for an actual outcome of the act of observing. Due to “macroscopic realism”, there can be only one actual outcome. Using the uniqueness quantifier of predicate logic, this is symbolically stated as\smallskip

\begin{equation}  %{Eq.85}
   \exists\mkern1mu\text{!}\mkern2mu{i}
   [\mkern-3.3mu[S_{i}(\text{``after''})]\mkern-3.3mu]
   =
   1
   \;\;\;\;  .
\end{equation}
\smallskip

\noindent Therefore,\smallskip

\begin{equation}  %{Eq.86}
   \sum_{i=1}^{N}
      \mkern2mu
      [\mkern-3.3mu[S_{i}(\text{``after''})]\mkern-3.3mu]
   =
   1
   \;\;\;\;  .
\end{equation}
\smallskip

\noindent If truth values of $S_{i}(\text{``before''})$ are determined and bivalent, then – according to negative construal of quantum superposition – it is not the case that there is only one true statement $S_{i}(\text{``before''})$. Symbolically this is expressed as negation of the uniqueness quantification:\smallskip

\begin{equation}  %{Eq.87}
   \neg
   \exists\mkern1mu\text{!}\mkern2mu{i}
   [\mkern-3.3mu[S_{i}(\text{``before''})]\mkern-3.3mu]
   =
   1
   \;\;\;\;  .
\end{equation}
\smallskip

\noindent In terms of the existential and universal quantifiers, the above is equivalent to\smallskip

\begin{equation}  %{Eq.88}
   \forall{i}
   [\mkern-3.3mu[S_{i}(\text{``before''})]\mkern-3.3mu]
   =
   0
   \mkern4mu
   \bigsqcup
   \mkern4mu
   \exists\mkern2mu{i}\mkern2mu\exists\mkern2mu{j}
   \Big(
      i
      \neq
      j
      \sqcap
      [\mkern-3.3mu[S_{i}(\text{``before''})]\mkern-3.3mu]
      =
      1
      \sqcap
      [\mkern-3.3mu[S_{j}(\text{``before''})]\mkern-3.3mu]
      =
      1
   \Big)
   \;\;\;\;  ,
\end{equation}
\smallskip

\noindent where $j\in\{1,\dots,N\}$. Consequently, either none of the statements $S_{i}(\text{``before''})$ is true (which cannot be acceptable since it is necessary that the act of observing produces some outputs), or at least two of $S_{i}(\text{``before''})$ are true. Accordingly,\smallskip

\begin{equation}  %{Eq.89}
   \sum_{i=1}^{N}
      \mkern2mu
      [\mkern-3.3mu[S_{i}(\text{``before''})]\mkern-3.3mu]
   =
   M
   \in
   \{2,\dots,N\}
   \;\;\;\;  ,
\end{equation}
\smallskip

\noindent where $M$ is \emph{the number of potential outcomes}.\\

\noindent With this result, the problem of definite outcomes can be reduced to the simple yet perplexing inequality:

\begin{equation} \label{INEQ} %{Eq.90}
   \sum_{i=1}^{N}
      \mkern2mu
      [\mkern-3.3mu[S_{i}(\text{``before''})]\mkern-3.3mu]
   \mkern2mu
   >
   \mkern2mu
   \sum_{i=1}^{N}
      \mkern2mu
      [\mkern-3.3mu[S_{i}(\text{``after''})]\mkern-3.3mu]
   \;\;\;\;  ,
\end{equation}
\smallskip

\noindent that is, $M>1$.\\

\noindent Again, what lies at the root of this inequality is the use of a standard logical system for the analysis and appraisal of $S_{i}(\text{``before''})$, the statements concerning yet-to-be performed observation.\\

\noindent By contrast, in the alternative model of reasoning,\smallskip

\begin{equation}  %{Eq.91}
   \forall{i}
   [\mkern-3.3mu[S_{i}(\text{``before''})]\mkern-3.3mu]
   =
   \textnormal{NULL}
   \;\;\;\;  .
\end{equation}
\smallskip

\noindent Recall that the sum of two natural numbers $a$ and $b$ can be defined by way of either the operation of union of two sets (such that $a+b=|{A}\cup{B}|$, where $a=|A|$ and $b=|B|$) or recursion (in a manner that $a+b^{+}=(a+b)^{+}$, where $n^{+}$ is the successor of $n\in\mathbb{N}$, i.e., the number following $n$, and the statement \text{\guillemotleft}$a+0=a$\text{\guillemotright} is considered to be true; consequently, $1+1=1+0^{+}=(1+0)^{+}=1^{+}=2\mkern1.5mu$) \cite{Enderton77}. Now take into account the facts:\\[-15pt]

\begin{enumerate}[a)]
    \item no set, including the empty set, has a cardinality of \textnormal{NULL},
    \item the statement \text{\guillemotleft}\textnormal{NULL}+0=\textnormal{NULL}\text{\guillemotright} is false.
\end{enumerate}

\noindent Because of those facts, the operation of addition cannot be defined on the set of $[\mkern-3.3mu[S_{i}(\text{``before''})]\mkern-3.3mu]$. This can be interpreted as evidence that \emph{potential outcomes of observation} – unlike their actual counterparts – \emph{have no number}, i.e., possess no property of the sum. Accordingly, in the alternative model of reasoning, the inequality $M>1$ is meaningless and so is the problem of definite outcomes.\\

\noindent Following from this, as long as the bare instrumentalist description of quantum mechanics is provided with the alternative model of reasoning, the problem of definite outcomes does not take place.\\

\bibliographystyle{References}

\end{document}